\documentclass[secnumarabic,amssymb, nobibnotes, preprint, aps, longbibliography]{revtex4-1}
\usepackage{color}
\usepackage[table]{xcolor}

\usepackage{amsmath}
\usepackage{graphicx}

\begin{document}
\title{Evidence for freezing of charge degrees of freedom across a critical point in CeCoIn$_5$}
\author{Nikola Maksimovic}
\altaffiliation{Contact for correspondence, nikola\_maksimovic@berkeley.edu or analytis@berkeley.edu}
\affiliation{Department of Physics, University of California, Berkeley, California 94720, USA}
\affiliation{Materials Science Division, Lawrence Berkeley National Laboratory, Berkeley, California 94720, USA}

\author{Vikram Nagarajan}
\affiliation{Department of Physics, University of California, Berkeley, California 94720, USA}
\affiliation{Materials Science Division, Lawrence Berkeley National Laboratory, Berkeley, California 94720, USA}

\author{Amanda Gong}
\affiliation{Department of Physics, University of California, Berkeley, California 94720, USA}
\affiliation{Materials Science Division, Lawrence Berkeley National Laboratory, Berkeley, California 94720, USA}

\author{Fanghui Wan}
\affiliation{Department of Physics, University of California, Berkeley, California 94720, USA}
\affiliation{Materials Science Division, Lawrence Berkeley National Laboratory, Berkeley, California 94720, USA}

\author{Stefano Faubel}
\affiliation{Department of Physics, University of California, Berkeley, California 94720, USA}
\affiliation{Materials Science Division, Lawrence Berkeley National Laboratory, Berkeley, California 94720, USA}

\author{Ian M. Hayes}
\affiliation{Department of Physics, University of California, Berkeley, California 94720, USA}
\affiliation{Materials Science Division, Lawrence Berkeley National Laboratory, Berkeley, California 94720, USA}

\author{Sooyoung Jang}
\affiliation{Department of Physics, University of California, Berkeley, California 94720, USA}
\affiliation{Materials Science Division, Lawrence Berkeley National Laboratory, Berkeley, California 94720, USA}

\author{Jan Rusz}
\affiliation{Department of Physics and Astronomy, Uppsala University, Box 516, S-75120 Uppsala, Sweden}

\author{Peter M. Oppeneer}
\affiliation{Department of Physics and Astronomy, Uppsala University, Box 516, S-75120 Uppsala, Sweden}

\author{Tessa Cookmeyer}
\affiliation{Department of Physics, University of California, Berkeley, California 94720, USA}

\author{Yochai Werman}
\affiliation{Department of Physics, University of California, Berkeley, California 94720, USA}

\author{Ehud Altman}
\affiliation{Department of Physics, University of California, Berkeley, California 94720, USA}

\author{James G. Analytis$^*$}
\affiliation{Department of Physics, University of California, Berkeley, California 94720, USA}
\affiliation{Materials Science Division, Lawrence Berkeley National Laboratory, Berkeley, California 94720, USA}

\begin{abstract}
The presence of a quantum critical point separating two distinct zero-temperature phases is thought to underlie the `strange' metal state of many high-temperature superconductors. The nature of this quantum critical point, as well as a description of the resulting strange metal, are central open problems in condensed matter physics. In large part, the controversy stems from the lack of a clear broken symmetry to characterize the critical phase transition, and this challenge is no clearer than in the example of the unconventional superconductor CeCoIn$_5$. Through Hall effect and Fermi surface measurements of CeCoIn$_5$, in comparison to ab initio calculations, we find evidence for a critical point that connects two Fermi surfaces with different volumes without apparent symmetry-breaking, indicating the presence of a transition that involves an abrupt localization of one sector of the charge degrees of freedom. We present a model for the anomalous electrical Hall resistivity of this material based on the conductivity of valence charge fluctuations.

\end{abstract}
\maketitle

%\section{Introduction}
CeCoIn$_5$ exhibits remarkably similar properties to high-temperature superconductors~\cite{Petrovic2001,Bianchi2003,Paglione2003,Nakajima2007,Settai2001,Kohori2001,Sidorov2002,Zhou2013,Stock2008,Tokiwa2013}, including signatures of an underlying quantum critical point (QCP) and a `strange' metallic phase extending to temperatures well above the superconducting transition temperature. In many of these materials, the identity of the putative QCP is unclear, and their behavior is difficult to reconcile with conventional theories of quantum criticality. For example, there is often no clear symmetry-breaking phase in proximity, or no clear evidence for fluctuations of a symmetry-breaking order parameter as would be expected of a conventional QCP. This has stimulated theoretical studies of unconventional QCPs that either weakly break symmetry~\cite{Varma2006,Lederer2017}, or break no symmetries at all~\cite{Senthil2004}. In this paper, we bring evidence that CeCoIn$_5$ is proximate to a QCP where the density of itinerant electrons (i.e. the Fermi volume) changes discontinuously, and apparently without symmetry-breaking.

At the microscopic level, heavy fermion materials including CeCoIn$_5$, are described by a Kondo lattice model, where a half filled $f$-electron valence shell from cerium contributes localized spin-$1/2$ moments that coexist with a sea of itinerant conduction electrons. In the conventional metallic ground state of a heavy fermion material, the $f$-electrons, in spite of being spatially localized, appear to become an integral part of the itinerant metal. In particular, they contribute their full share to the total Fermi volume as prescribed by Luttinger's theorem~\cite{Oshikawa2000}. This phenomenon occurs through the formation of Kondo singlet correlations between the local $f$ moments and the conduction electrons, which effectively hybridize the $f$ level with the conduction sea.

A long-standing challenge has been to characterize a QCP in which the $f$-electrons recover their localized character and withdraw from the itinerant Fermi volume. Superficially, the remaining Fermi volume without $f$-electrons is in apparent violation of Luttinger's theorem. The loss of Fermi volume is therefore conventionally accompanied by a transition to a (antiferromagnetic) spin-density wave state, whereby Luttinger's theorem is recovered in the appropriately folded Brillouin zone associated with translational symmetry breaking~\cite{Si2010}. Indeed, in almost all prominent heavy fermion materials where such an $f$-electron delocalization transition has been observed, it is accompanied by translational symmetry breaking~\cite{Paschen2004,Schroder2000,Gegenwart1998,Custers2003}. Without symmetry breaking, the only known way to reconcile Luttinger's theorem with localized $f$-electron charge is to form a fractionalized Fermi liquid~\cite{Senthil2004,Senthil2003}. In this theoretically predicted phase, the $f$-electron charge remains localized to the cerium site, while the spin excitations of the $f$ moments are itinerant and form a neutral Fermi surface~\cite{Senthil2004,coleman_theories_1999, gegenwart_quantum_2008}. In this paper, we present transport and quantum oscillation measurements of CeCoIn$_5$ with small levels of chemical substitution, and compare the experimental data to ab initio calculations. The results provide evidence that CeCoIn$_5$ is near an $f$-electron delocalization critical point induced by small levels of electron-doping. The apparent lack of symmetry breaking opens the possibility that a fractionalized phase is formed. Conductivity calculations in the context of the fractionalized Fermi liquid model are able to qualitatively capture the remarkable behavior of the experimentally measured electrical Hall coefficient, providing indirect evidence for an exotic quantum critical point associated with fractionalization of $f$-electrons.

%\section{Results}
\begin{figure*}[!htbp]
\centering
\includegraphics[scale=0.8]{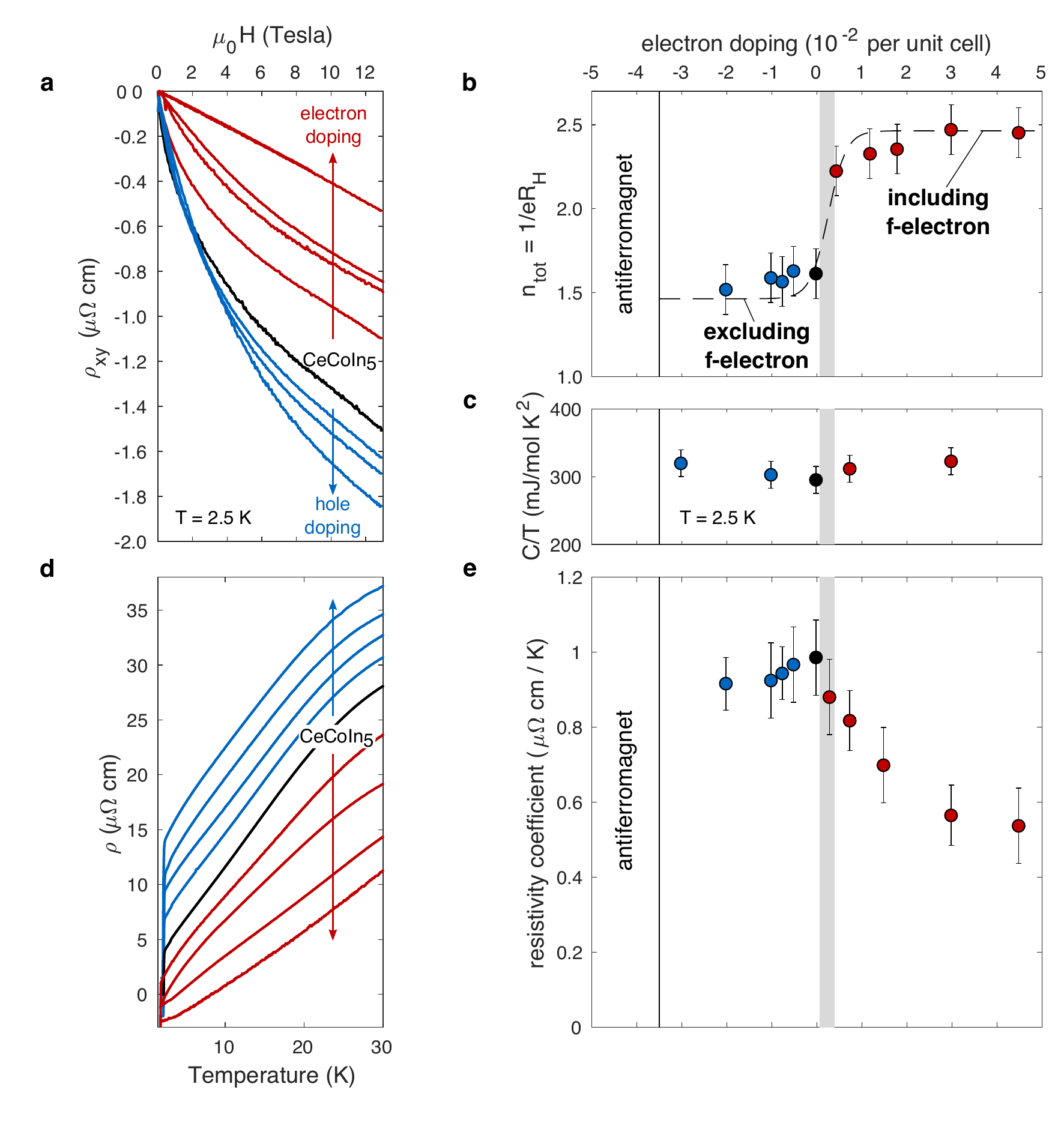}
\caption{\textbf{Transport and heat capacity of hole and electron doped CeCoIn$_5$} (a) Hall resistivity versus magnetic field at 2.5 Kelvin. (b) Net carrier density ($n_{tot}$) per unit cell, extracted from the linear slope of the Hall resistance between 11-13 Tesla at 2.5 K (or at 0.35 K shown in Supplement S9). $n_{\mathrm{tot}}$ exhibits a step when the material is electron-doped. The dashed black line indicates the expected $n_{\mathrm{tot}}$ including and excluding the $f$-electrons, as evaluated from measurements of LaCoIn$_5$ (see text). (c) Heat capacity coefficient at 2.5 Kelvin. (d) Resistance versus temperature; curves are offset vertically for clarity. (e) The slope of the resistance versus temperature (evaluated between 5 and 20 Kelvin) decreases rapidly when the material is electron-doped. Error bars are derived primarily from uncertainties in the measurements of geometric factors for transport samples, and sample masses for the heat capacity measurements.}
\label{fig:fig1}
\end{figure*}

The slope of the Hall resistivity at high fields can be used to estimate the net carrier density enclosed by the Fermi surface ($n_{\mathrm{tot}} = n_{\mathrm{holes}} - n_{\mathrm{electrons}}$)~\cite{Pippard2009}. Fig.\ 1a presents measurements of $\rho_{xy}$ vs $H$ for CeCoIn$_5$ samples with varying levels of cadmium or tin, both of which substitute indium. In many of these samples, the Hall resistivity is nonlinear at low fields. In this case, $n_{\mathrm{tot}}$ can be estimated using the slope of $\rho_{xy}$ vs $H$ in the high-field linear regime ($n_{\mathrm{tot}} = -\frac{1}{eR_{H}(H\rightarrow \infty)}$) (see Supplement S2). We extract an estimate of $n_{\mathrm{tot}}$ via the slope of $\rho_{xy}$ from 11-13 Tesla, or at lower fields in samples where $\rho_{xy}$ is completely $H$-linear. We can be confident that this method provides an accurate estimation of $n_{\mathrm{tot}}$ for two reasons. First, the value of $n_{\mathrm{tot}}$ in pure CeCoIn$_5$ extracted via this method at 2.5K agrees well with measurements at 50mK where $\rho_{xy}$ is completely linear in $H$~\cite{Singh2007}. Second, across this doping series the slope of $\rho_{xy}$ in the high-field linear regime is temperature-independent below 2.5K (Supplement S9). Over the same range, the nonlinearity of $\rho_{xy}$ at low fields is strongly suppressed by decreasing temperature. This suggests that the high-field slope of $\rho_{xy}$ below 2.5K is unaffected by temperature-dependent carrier mobilities, and is primarily determined by the temperature-independent carrier density.

Fig.~\ref{fig:fig1}b shows that $n_{\mathrm{tot}}$ of CeCoIn$_5$ extracted from these measurements is constant as a function of hole doping, but abruptly jumps to a higher density with slight electron doping. The carrier density of the conduction electrons excluding the $f$-electron can be established by measurements of the isostructural reference compound LaCoIn$_5$ (which can be thought of as CeCoIn$_5$ without $f$-electrons), which are shown in Supplement S7. We find that $n_{\mathrm{tot}}$ of CeCoIn$_5$ is close to that of LaCoIn$_5$, in agreement with previous literature~\cite{Singh2007}, suggesting that the $f$-electrons are not itinerant carriers in CeCoIn$_5$; in addition, $n_{\mathrm{tot}}$ remains constant with hole-doping within the experimental error. By contrast, slight electron-doping causes $n_{\mathrm{tot}}$ to jump by $1 \pm 0.2$ electrons per unit cell to a value consistent with the combined contributions of the conduction bands plus an itinerant $f$-electron per unit cell. This suggests that electron-doping drives the system through an $f$-electron delocalization QCP. There is no thermodynamic evidence for a finite temperature phase transition in the vicinity of this QCP (Supplement S3)~\cite{Pham2006}. Interestingly, the specific heat capacity at a moderate temperature remains constant across this doping series (Fig.~\ref{fig:fig1}c).

Fig.~\ref{fig:fig1}d shows that the longitudinal resistivity varies linearly with temperature, a hallmark of `strange' metal behavior, over an extended range in temperature across this doping series. Fig.~\ref{fig:fig1}e shows that the coefficient of the resistance versus temperature decreases rapidly as a function of electron-doping, highlighting the dramatic changes in the material's electronic properties upon Sn-substitution. While a description of strange metal resistivity is still not established, it is nevertheless thought that a factor of inverse carrier density enters the temperature coefficient of resistance in these materials~\cite{Bruin2013}, as it does in conventional metals. In this light, the nearly 60\% decrease in the temperature coefficient of the resistivity across this transition shown in Fig.~\ref{fig:fig1}e is consistent with a 70\% increase in carrier density found in Fig.~\ref{fig:fig1}b. While this is a natural interpretation, at present it is not possible to rule out the possibility that Sn-substitution also affects the $T$-dependent scattering rate.

\begin{figure*}[!htbp]
\centering
\includegraphics[scale=0.8]{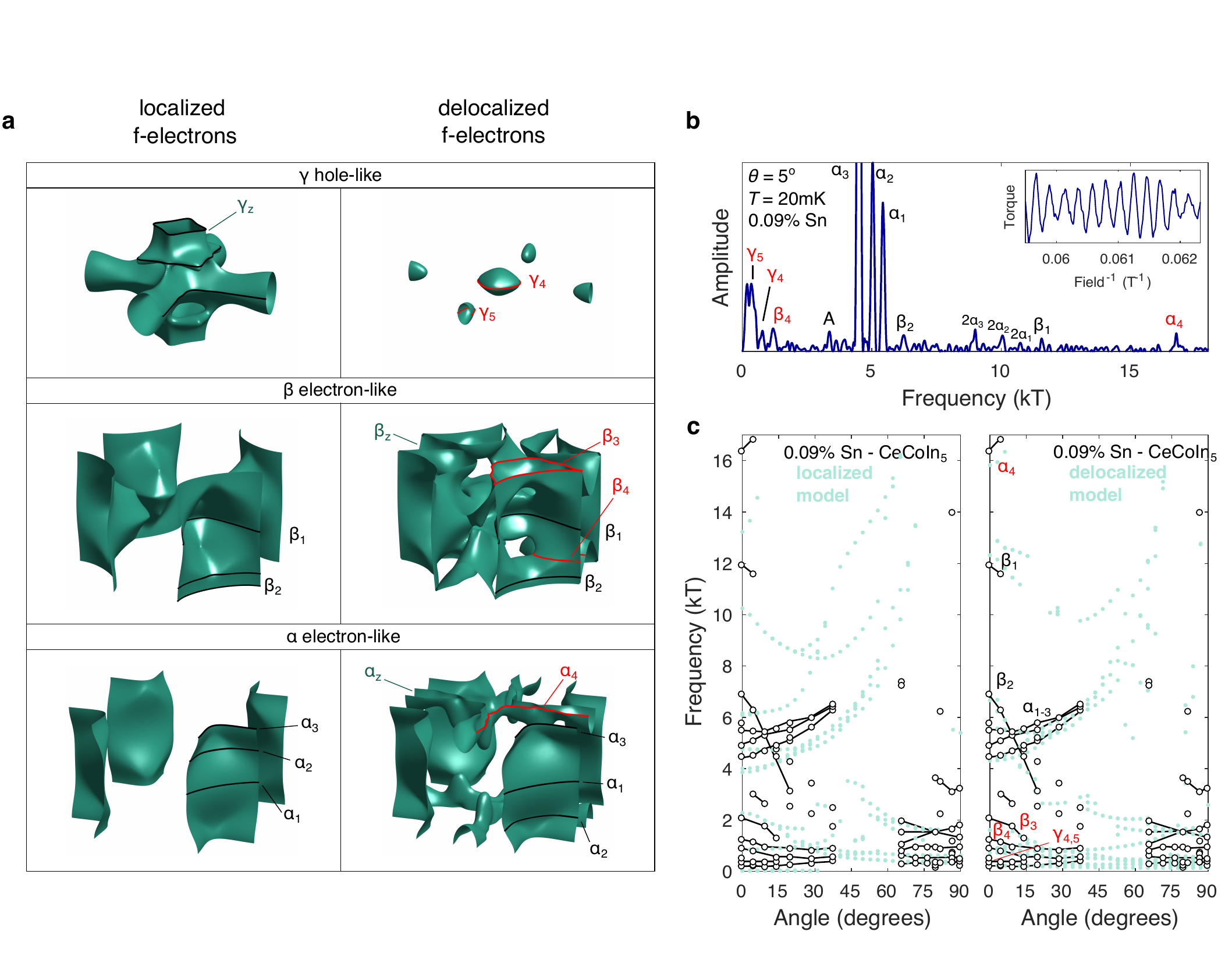}
\caption{\textbf{dHvA oscillations in 0.09\% Sn-substituted CeCoIn$_5$ and comparison to DFT calculations} (a) DFT calculated Fermi surface sheets of CeCoIn$_5$ with localized and delocalized $f$-electron models. Predicted dHvA orbits for $H \parallel$[001] are drawn in black and red. Red orbits are unique to the delocalized $f$-electron model. (b) Characteristic Fourier spectrum of dHvA oscillations ($\mu_{0}H$ = 13 - 17 T) with the magnetic field 5$^{o}$ away from [001] of a crystal of $\text{CeCo}(\text{In}_{0.9991}\text{Sn}_{0.0009})_{5}$. The inset shows raw oscillations after background subtraction. (c) Fundamental dHvA oscillation frequencies plotted as a function of angle tilting the magnetic field from the crystallographic [001] to [100] directions. Black points are taken from dHvA measurements of the 0.09\% Sn-substituted sample with black lines as guides to the eye. Light green points are DFT calculated frequencies of the localized and delocalized $f$-electron models respectively.}
\label{fig:QO}
\end{figure*}

When the $f$-electrons delocalize, the Fermi surfaces undergo a Fermi volume changing reconstruction. This is visualized in Fig.~\ref{fig:QO}a which shows results of our density functional theory (DFT) calculations of the three Fermi surfaces comparing the (de)localized $f$-electron models (DFT calculation details are provided in S1). $f$-electron delocalization causes the extended gamma surface to disconnect into small ellipsoidal pockets at the Brillouin zone center and edge, and the gamma pocket at the zone top ($\gamma_{Z}$) to disappear. Also, extended surfaces $\alpha_{Z}$ and $\beta_{Z}$ sheets appear at the zone top. In pure CeCoIn$_5$, ARPES reports stronger agreement with the localized $f$-electron model due to the absence of $\alpha_{Z}$ and $\beta_{Z}$~\cite{Chen2019,Chen2018,Jang2020}; the structure of $\gamma$ seems to be more controversial, being potentially disconnected but retaining $\gamma_{Z}$~\cite{Jang2020,Chen2018,Chen2019,Gauthier2020}. In addition, the volume of the $\alpha$ and $\beta$ cylinders in CeCoIn$_5$ increases slightly due to incipient $f$/conduction electron hybridization~\cite{Chen2019,Chen2018,Koitzsch2008}, suggesting that the $f$-electrons may have a weakly itinerant character. Such a nearly localized $f$-electron treatment is also promoted by magnetic resonance~\cite{Curro2001} and X-ray scattering data~\cite{Fujimori2016,Fujimori2003,Treske2014}, and is qualitatively consistent with the carrier density measurements in Fig.~\ref{fig:fig1}. 

Here we provide evidence that Sn-substitution induces a Fermi volume changing reconstruction consistent with full $f$-electron delocalization. De Haas-van Alphen (dHvA) oscillations measure extremal areas of the Fermi surface perpendicular to the field direction, giving a direct probe of the Fermi surface structure. Fig.~\ref{fig:QO}b shows a characteristic dHvA spectrum of a 0.9\% Sn-substituted sample of CeCoIn$_5$ (+0.5$e^{-}$/u.c.); spectra at other field-angles are provided in the Supplement. Although substitution of the indium site rapidly damps quantum oscillations, we find evidence for qualitative changes in the dHvA measurements compared to pristine CeCoIn$_5$, suggesting a Fermi surface reconstruction occurs. A comparison of the model calculations and experimental data for H$\parallel$[001] is tabulated in Table~\ref{table1}, and a full angle-dependent map is shown in Fig.~\ref{fig:QO}c. First, a new frequency $\alpha_{4}$ at about 16kT for H near [001] agrees well with a predicted orbit on $\alpha_{Z}$ of the itinerant model. This suggests that the $\alpha_{Z}$ Fermi surface emerges in the Sn-substituted sample. Second, $\beta_{2}$ decreases as a function of tilt angle in the Sn-substituted sample in better agreement with the itinerant model due to the presence of the $\beta_{Z}$ Fermi surface. This is in contrast to pure CeCoIn$_5$, where $\beta_{2}$ increases as a function of tilt angle~\cite{Settai2001} reflecting the absence of $\beta_{Z}$. Accordingly, in the Sn-substituted sample we assign the 1.2kT and 2.0kT frequencies for H$\parallel$[001] to orbits on $\beta_{Z}$; the angle-dependence of these orbits agrees well with those of the itinerant model calculations (Fig.~\ref{fig:QO}c). Taken together, these features suggest that $\beta_{Z}$ emerges in the Sn-substituted sample. Finally, there are a number of persistent $<1$kT orbits (more clearly visible in spectra shown in the Supplement). These could be orbits on disconnected $\gamma$ surface ellipsoids of the itinerant model, but their origin is uncertain due to the fact that both models have a number of orbits $<1$kT; CeCoIn$_5$ also shows low-frequency orbits with unknown origin at particular angles between [001] and [100]~\cite{Settai2001}. Nevertheless, the appearance of $\alpha_{4}$, $\beta_{3}$ and $\beta_{4}$, and the change in slope of $\beta_{2}$ as a function of tilt angle indicate that Sn-substitution of CeCoIn$_5$ induces a Fermi surface reconstruction associated with the appearance of $\alpha_{Z}$ and $\beta_{Z}$ --- relatively large Fermi surfaces of the itinerant $f$-electron model which are not detected in the pure compound. The comparison of dHvA data and DFT calculations, from the perspective of the measured Fermi surface, corroborates the evidence in Fig.~\ref{fig:fig1} for an $f$-electron delocalization transition.

\begin{table}[]
\begin{center}
\begin{tabular}{|c|c|c|c|c|c|}
\hline
\begin{tabular}[c]{@{}l@{}} Fermi surface
\end{tabular} 
&\begin{tabular}[c]{@{}l@{}}dHvA 
\\orbit 
\\label
\end{tabular}
&\begin{tabular}[c]{@{}l@{}}localized $f$-electron\\
model
\end{tabular}
&\begin{tabular}[c]{@{}l@{}}CeCoIn$_5$\\ 
Ref.~\cite{Settai2001}
\end{tabular}
&\begin{tabular}[c]{@{}l@{}}0.09\% Sn-doped CeCoIn$_5$ 
\end{tabular} 
&\begin{tabular}[c]{@{}l@{}}delocalized $f$-electron\\ \, \; \; \; model\\ \end{tabular}
\\ 
\hline \hline
$\gamma_{Z}$ & $\gamma_{1}$& 0.8 & & & \\
$\gamma_{Z}$ & $\gamma_{2}$& 2.3 & & & \\
$\gamma$-cross & $\gamma_{3}$& 13.2 & & & \\ 
$\gamma$-ellipsoid & $\gamma_{4}$& & & (0.86) & 0.7 \\
$\gamma$-ellipsoid & $\gamma_{5}$& & & (0.17) & 0.22 \\
\hline
$\alpha$-cylinder & $\alpha_{1}$& 4.8 & 5.6 & 5.4 & 5.6\\
$\alpha$-cylinder & $\alpha_{2}$& 4.0 & 4.5 & 4.8 & 4.4\\
$\alpha$-cylinder & $\alpha_{3}$& 3.9 & 4.2 & 4.4 & 4.3\\ 
$\alpha_{Z}$ & $\alpha_{4}$& & & 16.3 & 15.8\\
\hline
$\beta$-cylinder & $\beta_{1}$& 10.3 & 12.0 & 11.9 & 12.3\\
$\beta$-cylinder & $\beta_{2}$& 6.1 & 7.5 &  6.8 & 6.7\\ 
$\beta_{Z}$ & $\beta_{3}$& & & 2.0 & 1.6\\
$\beta_{Z}$ & $\beta_{4}$& & & 1.2 & 0.9\\
\hline
\end{tabular}
\caption{de Haas-van Alphen extremal orbit assignments (units of kiloTesla, H $\parallel$ [001]) from experiments and DFT calculations. Each orbit is labeled by the corresponding Fermi surface pocket, which are visualized on the calculated Fermi surface sheets in Fig.~\ref{fig:QO}a.}
\label{table1}
\end{center}
\end{table}

\begin{figure}[!htbp]
\centering
\includegraphics[scale=0.8]{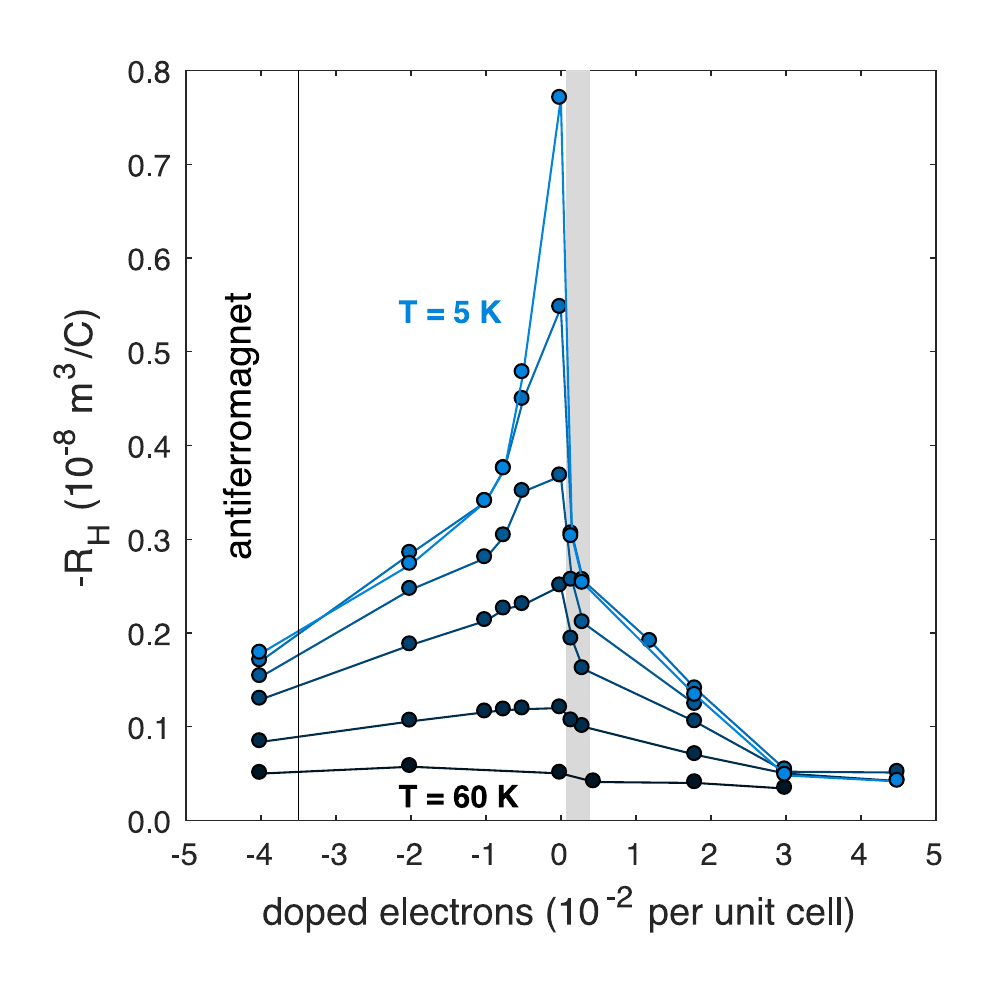}
\caption{\textbf{Anomalous enhancement of low-field Hall coefficient} extracted from $R_{H} = \rho_{xy}/\mu_{0}H$ at an applied field of $\mu_{0}H =$ 0.1T taken at various temperatures (T = 5 K, 10 K, 15 K, 20 K, 30 K, 60 K).}
\label{fig:fig3_diverging}
\end{figure}

Fig.~\ref{fig:fig3_diverging} shows that the Hall coefficient is massively enhanced across this substitution series when the external magnetic field is reduced to zero. The symmetric enhancement of the low-field Hall coefficient as observed in Fig.~\ref{fig:fig3_diverging} across the critical point is remarkable. Conventionally, the low-field limit of the Hall effect is inversely proportional to the carrier density of the most mobile carriers~\cite{Pippard2009}. It is therefore unusual for the low-field Hall coefficient to move in the same direction with either hole or electron doping, even in the case of a band structure singularity. This symmetric-in-doping Hall coefficient cannot be attributed to disorder scattering induced by substitution, as we find that disordering the material by other means, substituting lanthanum for cerium, has essentially no effect on the measured Hall coefficient (see Supplement S5). Having established evidence for an $f$-electron delocalization QCP, we pursue theoretical explanations of the unconventional Hall effect in the context of $f$-electron delocalization and valence fluctuation phenomena.

\begin{figure}[!htbp]
\centering
\includegraphics[scale=0.8]{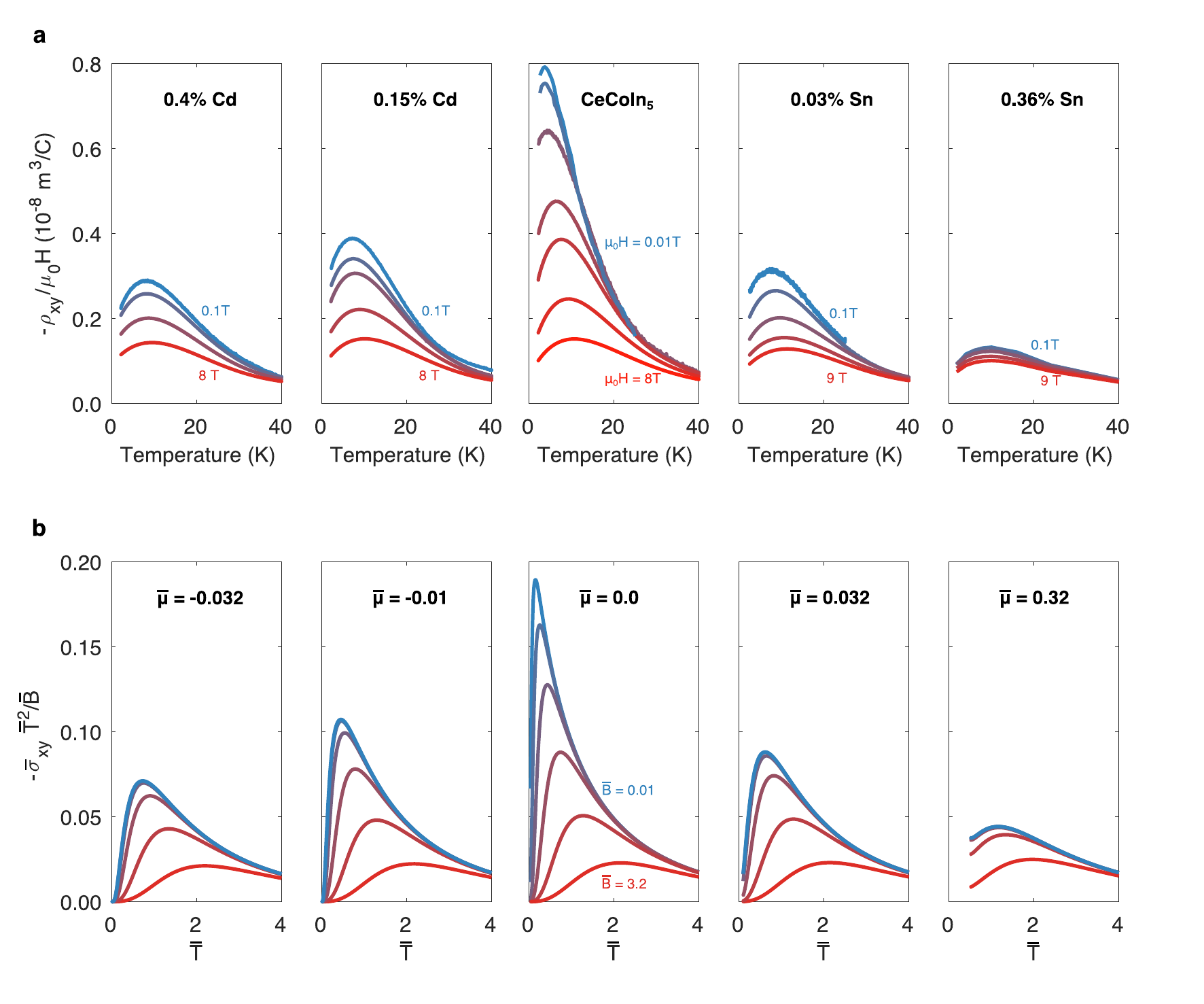}
\caption{\textbf{Comparison of experimental data and theoretical calculations of the conductivity of critical valence fluctuations} (a) Experimentally measured Hall resistivity, divided by the applied magnetic field, for samples with different compositions. Different traces are taken at different applied magnetic fields (0.1, 1, 3, 6, 9 T). Each panel is labeled by the substitution level. (b) The theoretically predicted Hall effect due to bosonic valence fluctuations of the fractionalized Fermi liquid model. Each panel is labeled by the chemical potential in the theory corresponding to the doping level in the experiment, where $\mu < 0$ corresponds to hole-doping and $\mu > 0$ corresponds to electron-doping. Curves labeled by the normalized magnetic field value ($\bar{B}$ = 0.01, 0.032, 0.1, 0.32, 1, 3.2) and all theory data includes a parametrization of impurity scattering, $\bar C=4$. See Supplement S6 for the details of the calculation and relevant parameter normalizations.}
\label{fig:fig3}
\end{figure}

The simplest mechanism for the $f$-electron delocalization transition is the increase (decrease) in `Kondo' coupling between the $f$ moments and the conduction electrons induced by electron (hole) doping ~\cite{Gofryk2012,Chen2018,Sakai2015}. When the Kondo coupling is sufficiently strong, singlet bonds form between the $f$ level and the conduction electrons, stabilizing a Fermi liquid with a Fermi volume that includes the $f$-electrons. Due to constraints imposed by Luttinger's theorem~\cite{Oshikawa2000}, the loss of Fermi volume when the $f$-electrons localize is expected to coincide with an antiferromagnetic phase where the Brillouin zone is reduced~\cite{Si2010}. It is however hard to reconcile this scenario with the data in this work because the transition to antiferromagnetism is seen only around Cd doping of $0.6\%$~\cite{Pham2006}, considerably removed from the suggested delocalization transition induced by Sn-substitution. Furthermore, magnetic order has never been observed in CeCoIn$_5$ or Sn-substituted CeCoIn$_5$ despite intensive efforts~\cite{Sakai2015,Chen2018,Gofryk2012,Kohori2001}, and our dHvA data suggests that the Brillouin zone is essentially unchanged by slight Sn-substitution. We also note that the magnetic fields applied in this report are likely too small to affect the Kondo coupling scale, which has been suggested to be on the order of 50K in this material~\cite{Jang2020}. Indeed, we do not detect nonlinearity in magnetic susceptibility or field-dependent Fermi volume (see Supplement S3 and S8), suggesting that the data taken at high-field are representative of the zero-field condition. Therefore, the lack of symmetry breaking across this doping series opens the possibility for a fractionalized phase in the localized $f$-electron regime. In this phase, the $f$-electrons decouple from the conduction band to form a gapless spin liquid with a neutral Fermi surface which coexists with the conduction electrons~\cite{Senthil2004}. In light of this possibility, we can speculate that the specific heat remains constant across the substitution series (Fig.~\ref{fig:fig1}b,c) due to the presence of a neutral Fermi surface, which preserves the specific heat of the system even when the electrical carrier density appears to decrease in the localized $f$-electron phase. Indeed, the specific heat coefficient of CeCoIn$_5$ is known to be anomalously large when compared to the measured Fermi volume~\cite{Settai2001}, suggesting there may be charge-neutral contributions to the low-temperature heat capacity.

Fractionalized excitations have proven difficult to measure in experiments even in well-established cases, and often indirect evidence is used. Here we show that a model of fractionalized charge and spin can account for the remarkable enhancement of the Hall coefficient seen in the experiment in Fig.~\ref{fig:fig3_diverging}. In the simplest description of the fractionalized Fermi liquid, the $f$-electron separates into a fermionic spinon carrying its spin, and a gapped bosonic mode, in this case representing a valence charge fluctuation. At finite temperature, the electrical conductivity has contributions from the fermionic spinons, the charged bosons and the conduction electrons. The spinon and the bosons should be added in series~\cite{Ioffe1989}. The boson's resistivity will then dominate due to their much smaller number, and we therefore neglect the spinon contribution. Adding to this the resistivity of the conduction band in parallel gives:

\begin{equation}
    R_H= R_H^c {\sigma_c^2\over (\sigma_{\text{tot}})^2}+\frac{1}{\mu_{0}H} {\sigma_{xy}^b\over (\sigma_{\text{tot}})^2} 
    \label{eq:R_H}
\end{equation}
where $\sigma_c$ and $R_H^c$ are the longitudinal conductivity and Hall coefficient of the conduction electrons, respectively, and $\sigma_{xy}^b$ is the Hall conductivity of the critical valence fluctuations. The total conductivity is $\sigma_{\text{tot}}$. In our calculation, we consider two processes that contribute to the scattering rate of the valence fluctuations. One process is provided by the internal gauge field~\cite{Senthil2004}. The other mechanism is scattering on the doped ions, which grows linearly with the doping level (see Supplement S5). One may expect an enhancement of the Hall coefficient stemming from the second term in Eq.~\ref{eq:R_H} due to the singular behavior of the valence fluctuations at the critical point. This expectation is corroborated by a semi-classical Boltzmann analysis, the details of which are given in Supplement S6. As seen in Fig. \ref{fig:fig3}, the results of the calculation of the conductivity in this model give good agreement with the measured Hall coefficient across the doping series. The results shown in Fig.\ \ref{fig:fig3}b are obtained from a calculation of $\sigma^b_{xy}$, and converted to a Hall coefficient using the physical resistivity of the system $1/\sigma_{\text{tot}} = \rho_{xx} \sim T$ as observed in the experiment over the relevant temperature range. It remains to be seen what sort of scattering processes produce the linear-in-temperature longitudinal resistivity of CeCoIn$_5$. The addition of Boltzmann processes is not expected to qualitatively affect our calculation of the Hall coefficient, which is dominated by the contribution of critical valence fluctuations.

The present study provides evidence that CeCoIn$_5$ exists near a quantum critical point associated with the delocalization of $f$-electron charge. The lack of symmetry breaking around this transition opens the possibility for the formation of a fractionalized Fermi liquid in the localized $f$-electron phase. While the consistency of the Hall data with our transport calculations in this framework support this theoretical picture, direct evidence for fermionic magnetic excitations and spinless bosonic charge fluctuations would be desireable. This may be possible using inelastic neutron measurements~\cite{Banerjee2018} or Josephson tunneling experiments~\cite{Senthil2001}. On a final note, an increasingly popular hypothesis for the underlying QCP of the cuprate high-temperature superconductors is a Fermi surface reconstruction where localized moments, in that case of a Mott insulator, become itinerant (known as a $p$ to $1+p$ transition~\cite{Badoux2016}). We have presented evidence for an analogous transition in a Kondo lattice, where the localized charge of the $f$-electrons becomes itinerant. It is possible that such a QCP underlies some of the striking similarities between CeCoIn$_5$ and cuprate superconductors~\cite{Paglione2003,Nakajima2007,Settai2001,Petrovic2001,Kohori2001,Sidorov2002,Zhou2013,Stock2008,Bianchi2003}.

\section{Acknowledgements}
We would like the thank C. Varma, R. McDonald, S. Sachdev, S. Chatterjee, M. Vojta, and J.D. Denlinger for helpful discussions. We thank E. Green and A. Bangura for support during experiments at the milliKelvin facility in the National High Magnetic Field Lab. V. N. is supported by the National Science Foundation Graduate Research Fellowship Grant No. DGE-1752814. This work is supported by the Gordon and Betty Moore Foundations EPiQS Initiative through Grant GBMF9067. P. M. O. and J. R. are supported by the Swedish Research Council (VR), and the K. and A. Wallenberg Foundation Award No. 2015.0060. DFT calculations have been performed using resources of Swedish National Infrastructure for Computing (SNIC) at the NSC center (cluster Tetralith). dHvA measurements were performed at the National High Magnetic Field Laboratory, which is supported by the National Science Foundation Cooperative Agreement No. DMR-1644779 and the State of Florida. 

\section{Author contributions}
N.M. and J.G.A conceived of the experiment. N.M and I.M.H. performed the Hall effect measurements. N.M. and V.N. performed the quantum oscillation experiments. N.M., S.F., F.G. and A.G. grew the samples. T.C., Y.W., and E.A. performed theoretical calculations of the Hall coefficient. J.R. and P.M.O. performed DFT simulations of Fermi surface topologies and dHvA oscillation frequencies. All authors contributed to writing the manuscript.

\section{Data availability}
All data provided in this report are publicly available at https://osf.io/dfm7x/.

%\bibliography{references}

\newpage
\section*{Supplement for ``Evidence for freezing of charge degrees of freedom across a critical point in CeCoIn$_5$''}
\section*{S1 Methods}
Single crystals of $\text{CeCoIn}_{5}$ were grown by an indium self-flux described elsewhere with a nominal concentration of indium flux replaced by cadmium or tin~\cite{Petrovic2001,Pham2006}. Hall bar devices were prepared by mechanically thinning single crystals along the crystallographic $c$-axis to $<$20 $\mu\text{m}$ thickness and attaching gold wires with EpoTek EE129 on gold-sputtered pads. Geometric factors were measured with an optical microscope, and crystals with different doping levels were thinned simultaneously to eliminate statistical error in the thickness measurement. Transport in the $ab$-plane was measured using the standard lockin technique with current excitations of 1-3 mA and magnetic field directed along the crystallographic $c$-axis in a 14T Quantum Design PPMS. Hall resistance was anti-symmetrized with respect to field polarity. Volume magnetization was measured with a SQUID magnetometer. Heat capacity was measured in a QuantumDesign PPMS.

de Haas-van Alphen experiments were carried out at the milliKelvin facility at the National High Magnetic Field lab in Tallahassee, Florida. The sample was mounted with silicon grease on a piezoelectric torque cantilever (120 $\mu$m length). The deflection of the cantilever in an external magnetic field was measured through a Wheatstone bridge and a Lock-in amplifier with a 100 $\mu$A source current.

Samples are labeled by the measured concentration as determined by microprobe analysis. Systematic increases in doping concentration for each species were confirmed by a combination of microprobe measurements and, where applicable, a resistive measurement of the superconducting transition temperature in accordance with established literature values~\cite{Pham2006,Chen2018}. Because of the low concentrations used in this report, we applied microprobe analysis to higher doping levels for accuracy, and extrapolated the measured linear dependence of true concentration versus nominal concentration. Using this method, the true concentration of dopants was found to be lower than the nominal concentration in the flux, agreeing with previous Cd (10\%) and Sn (60\%) alloying studies on $\text{CeCoIn}_{5}$~\cite{Pham2006}.

Density functional theory (DFT) calculations have been performed using the full-potential linearized augmented plane waves (FP-LAPW) method, as implemented in the WIEN2k code~\cite{Wein2K_1}. For the self-consistent cycle we have used 40000 k-points in the full Brillouin zone (2520 k-points within its irreducible wedge), basis size was over 850, determined by the $RK_\mathrm{max}=8.0$ parameter. Spin-orbital interaction was included within a second-variational treatment~\cite{Kunes2001} using a basis of approximately 600 scalar-relativistic eigenfunctions (parameter $E_\mathrm{max}=5.0$~Ry). Exchange-correlation effects were treated within local density approximation (LDA)~\cite{Perdew1992}. Lattice parameters of CeCoIn$_5$ were set equally as in the Ref.~\cite{Elgazzar2004}.

The de Haas-van Alphen (dHvA) frequencies were calculated using SKEAF code developed by Rourke and Julian~\cite{Rourke2012}. For this purpose a finer mesh of k-points has been generated, consisting of 41106 k-points in the irreducible wedge of the Brillouin zone (sampling the whole Brillouin zone by 100x100x60 k-points). Resulting band-structure was interpolated by SKEAF using 200 grid points per single side, resulting in well converged extremal orbits. All other parameters were kept at default values. For evaluating the angular dependence of dHvA frequencies we have used 30 rotation steps. Our calculated dHvA frequencies for $H || c$ are in excellent agreement with earlier calculations~\cite{Elgazzar2004}. Note however the appearance of a large 16~kT extremal orbit from the $\alpha$ Fermi surface sheet, which originates from only a minute difference in the calculated band structures. The difference is likely caused by different parametrizations of LDA --- Perdew \& Wang (1992)~\cite{Perdew1992} used in the present work vs Perdew \& Zunger (1981) used in Ref.~\cite{Elgazzar2004}.
\newpage

\section*{S2 Hall effect in doped CeCoIn$_5$}

\begin{figure*}[!htbp]
\centering
\includegraphics[scale=0.8]{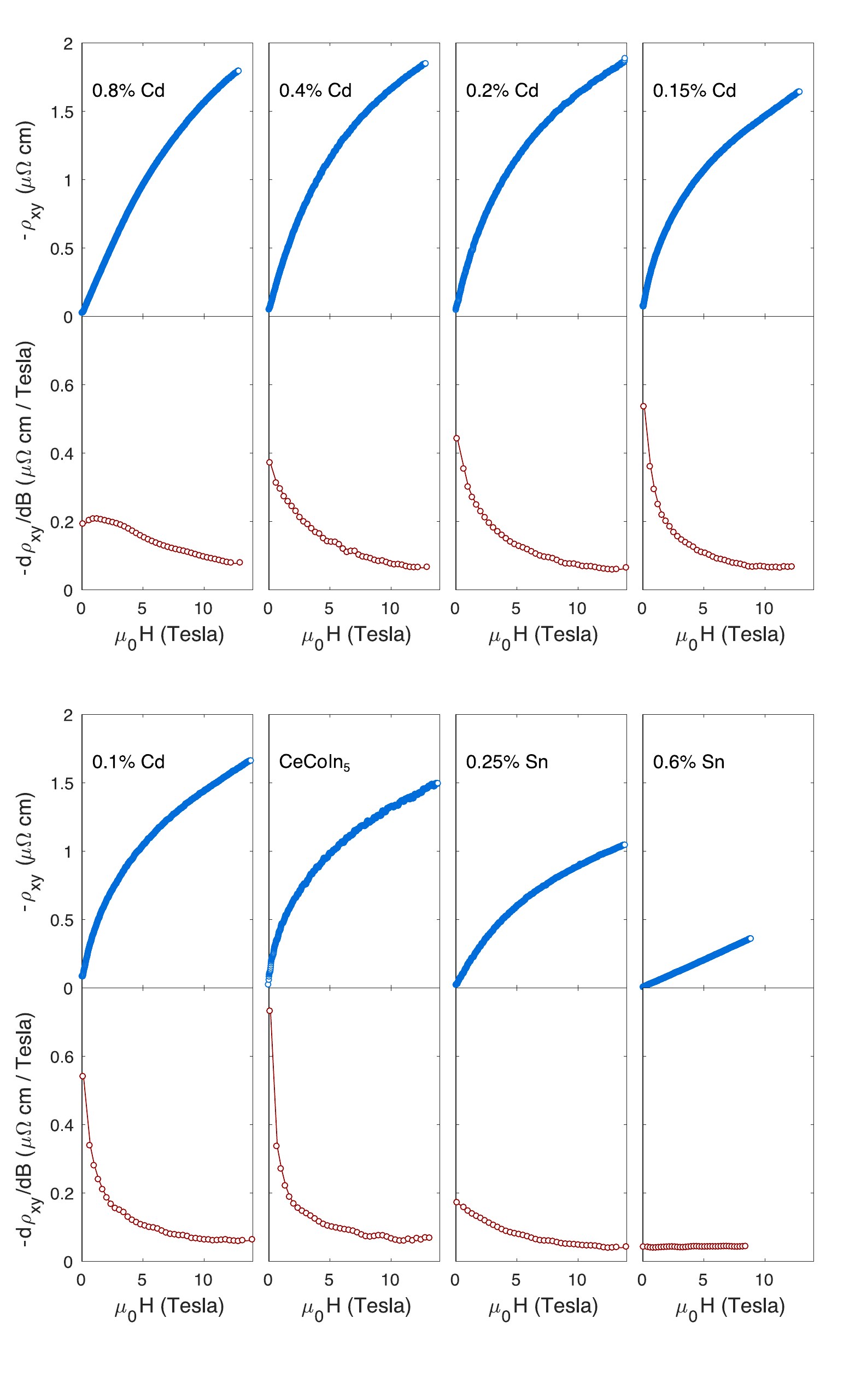}
\caption{\textbf{Hall resistivity versus field at T = 4K} Blue curves are measured Hall resistivity, and red curves below each panel are the derivative with respect to field. In all samples, the Hall resistivity approaches a linear dependence at high field.}
\label{fig:sup_hallderiv}
\end{figure*}

Fig.~\ref{fig:sup_hallderiv} shows the Hall resistivity at T = 4K for samples with different substitution levels. In all samples, the effective slope of the Hall resistivity versus field $R_{H} \sim d\rho_{xy}/dB$ approaches a constant above about 11 Tesla. From the value of $d\rho_{xy}/dB$ at high fields, a net carrier density can be extracted using the standard formula $n_{\text{tot}} = n_{h} - n_{e} = -\frac{1}{eR_{H}}$~\cite{Pippard2009}. This value at high fields (approximately 0.061 m$^{3}$/C) in CeCoIn$_5$ and Cd-doped samples corresponds to a net electron-like carrier density about 10\% larger than that of LaCoIn$_5$. Examining the Hall resistivity across the doping range, there are a few notable trends. First, the 0.8\% Cd sample is the only one that shows antiferromagnetic order at low temperatures. The slope of the Hall resistivity versus field shows a non-monotonic dependence in this sample. When AFM is destroyed by lowering the Cd concentration, the Hall coefficient has a pronounced upturn at zero field. This upturn becomes more pronounced when approaching CeCoIn$_5$ from the Cd-doped side, and is strongest in pristine CeCoIn$_5$. Upon substituting with tin (electron-doping), the low-field divergence of $\rho_{xy}$ becomes suppressed.

Note also the highly nonlinear Hall resistivity as a function of magnetic field. In a 0.6\% Sn substituted sample, this curvature is absent, and Hall resistivity is completely linear in field. 

\begin{figure*}[!htbp]
\centering
\includegraphics[scale=0.8]{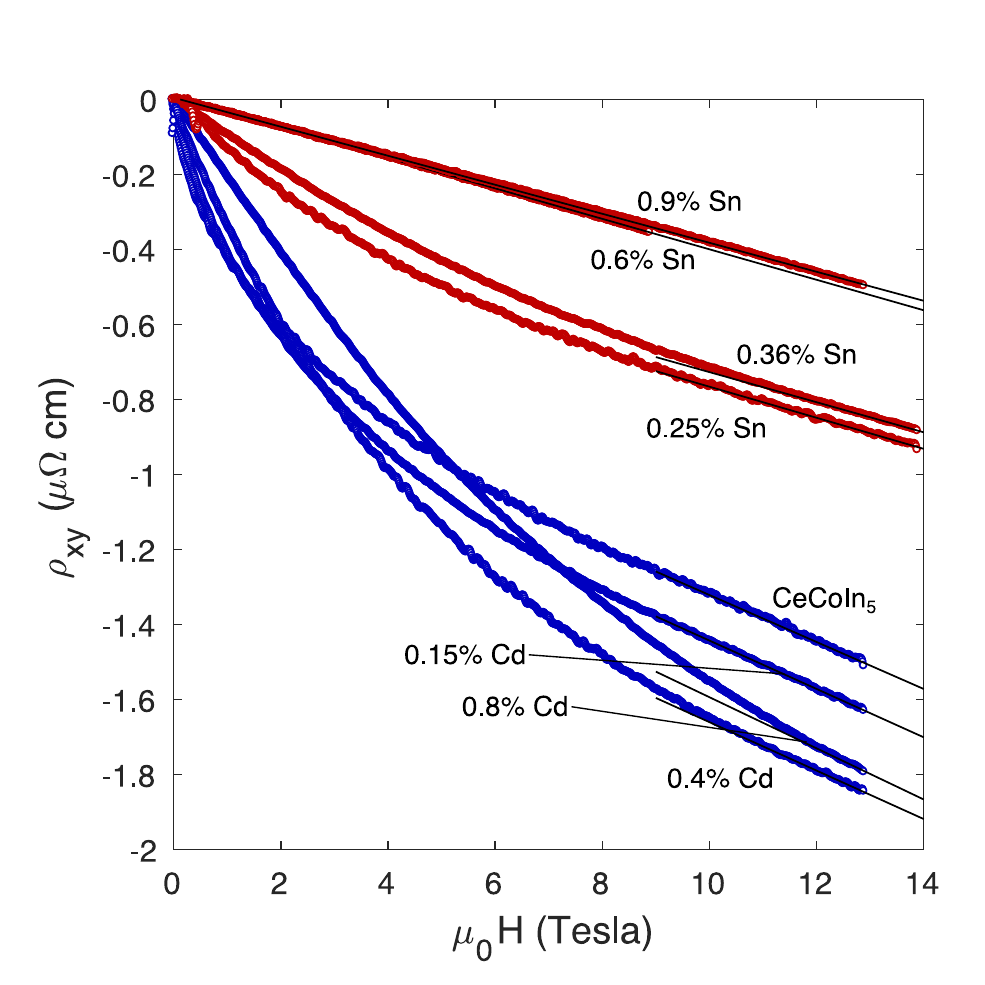}
\caption{\textbf{Hall resistivity versus field at T = 2.5K} Blue curves are Hall resistivity traces of cadmium doped samples. Red curves are traces of Sn doped samples. The solid lines show extractions of the high-field slope.}
\label{fig:sup_Hall_analysis}
\end{figure*}

Fig.~\ref{fig:sup_Hall_analysis} shows traces of the isothermal Hall resistance against applied magnetic field for several dopings. From each of these traces, a high-field Hall slope (corresponding to a Hall coefficient) is extracted and converted to a carrier density presented in Fig.~\ref{fig:fig1}c of the manuscript. Fig.~\ref{fig:sup_Hall_analysis} also highlights the qualitative difference in the behavior of Sn-substituted samples and Cd-substituted samples. For example, the Hall slope is completely linear in more strongly Sn-substituted samples.

\newpage
\section*{S3 Magnetization}
\begin{figure*}[!htbp]
\centering
\includegraphics[scale=0.8]{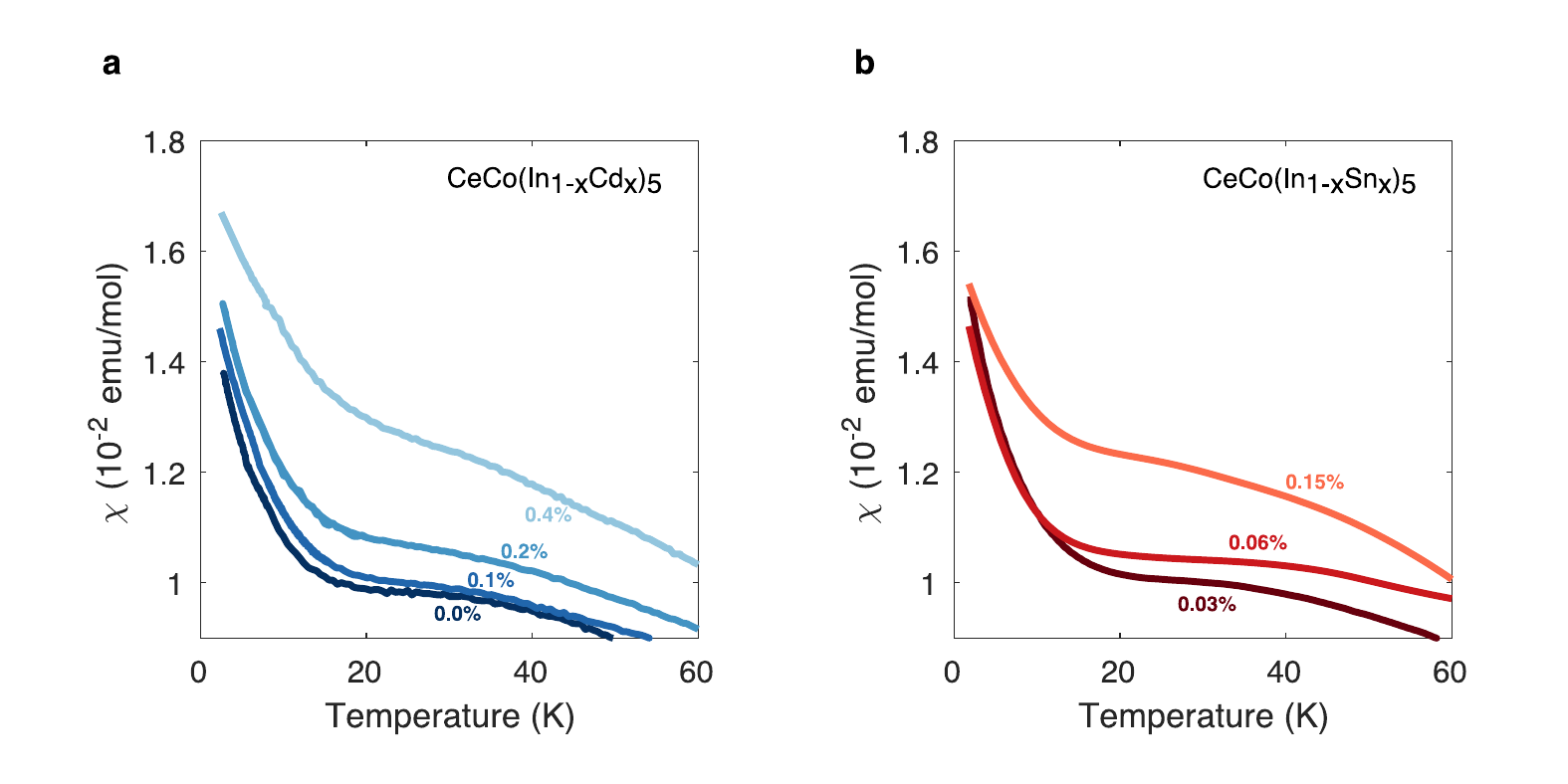}
\caption{\textbf{c-axis magnetic susceptibility versus temperature} Measured for varying doping levels of cadmium and tin ($\mu_{0}H$ = 0.1T applied along the c-axis) in the zero-field cooled condition. The superconducting transition was truncated for clarity.}
\label{fig:sup_magvB}
\end{figure*}

\begin{figure*}[!htbp]
\centering
\includegraphics[scale=0.8]{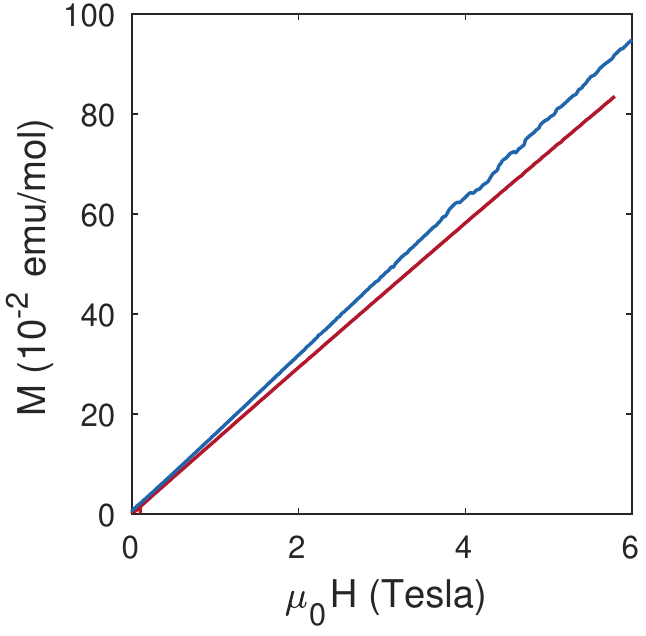}
\caption{\textbf{c-axis magnetic susceptibility versus field at 10K} In CeCoIn$_5$ substituted with 0.2\% Cd (blue line), and 0.3\% Sn (red line). The magnetization is linear-in-field up to 6 Tesla.}
\label{fig:sup_mag}
\end{figure*}

Here we present magnetization data for samples with different doping levels. Fig.~\ref{fig:sup_mag} shows that there is no evidence for a phase transition in the temperature-dependent magnetic susceptibility above the superconducting transition. This is consistent with the thermodynamic dataset in Ref.~\cite{Howald2015}. Fig.~\ref{fig:sup_magvB} shows that the magnetization is completely linear up to 6 Tesla in both Cd-substituted and Sn-substituted samples at low temperature.

\newpage
\section*{S4 Heat capacity measurements}
\begin{figure*}[!htbp]
\centering
\includegraphics[scale=0.8]{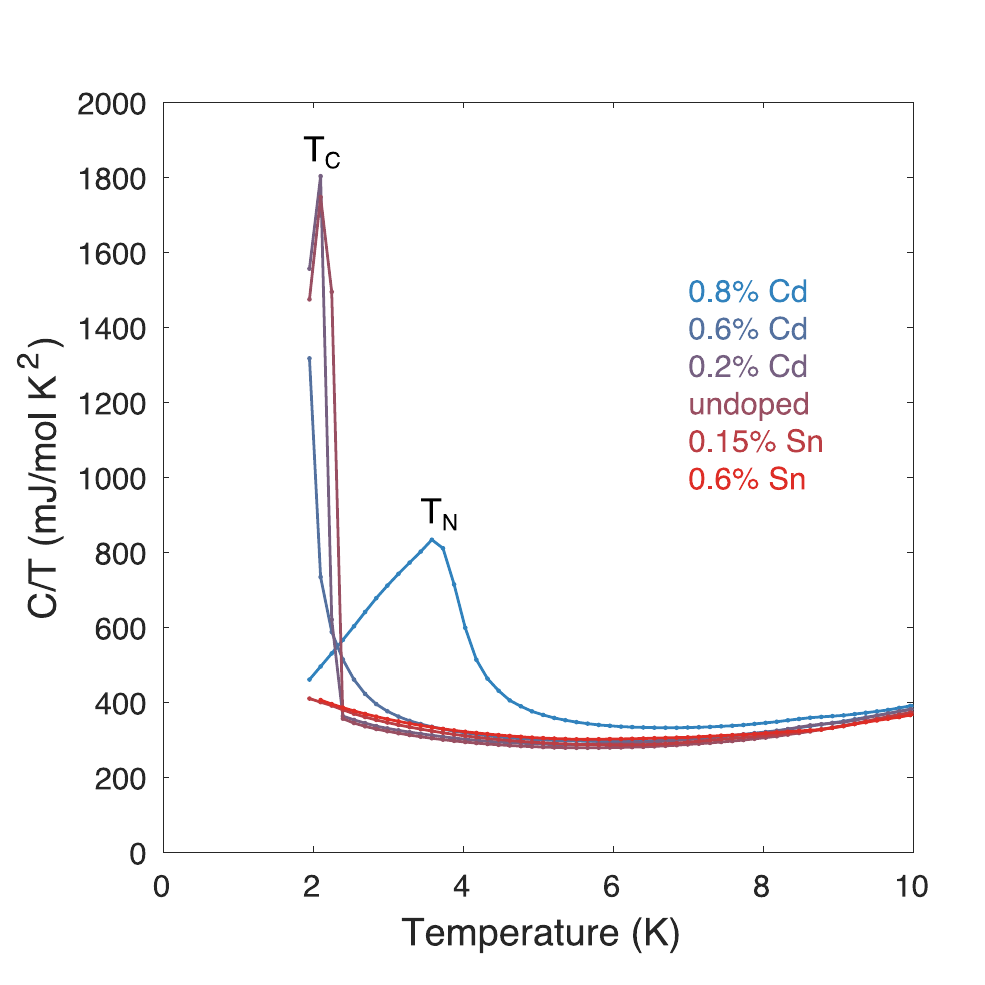}
\caption{\textbf{Heat capacity} Specific heat capacity for a variety of compositions of CeCoIn$_5$ doped with either cadmium or tin on the indium site. Peaks in the heat capacity correspond to either the superconducting transition $T_{c}$, or the AFM transition $T_{N} = 3.6$K, which only appears in the 8\% Cd substituted sample.}
\label{fig:sup_HC}
\end{figure*}

Heat capacity gives a measure of the number of entropy-carrying degrees of freedom. Fig.~\ref{fig:sup_HC} shows the heat capacity plotted as $C/T$ vs $T$. This material is known to have an anomalously temperature-dependent heat capacity coefficient~\cite{Petrovic2001,Bianchi2003}. For simplicity, in the main text we compare the heat capacity between different samples at a fixed temperature, though the qualitative results are not temperature-dependent. The heat capacity is largely unaffected by the doping levels used in this report. Superconducting and/or AFM transitions are visible in heat capacity measurements down to 1.8K.

\newpage

\section*{S5 Site-dependent effects of doping}
\begin{figure*}[!htbp]
\centering
\includegraphics[scale=0.8]{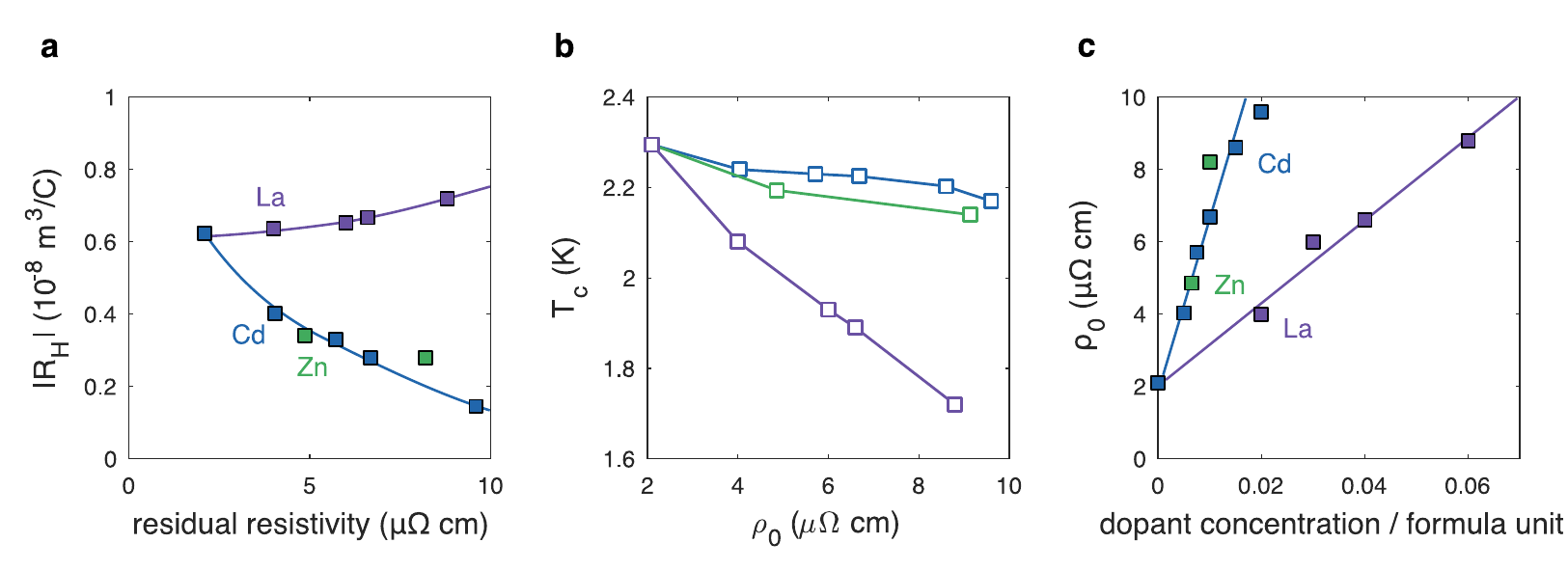}
\caption{\textbf{Effect of different hole-dopant species on transport quantities} Lanthanum substitution of the cerium site (purple), cadmium substitution of the indium site (blue), and zinc substitution of the indium site (green) in CeCoIn$_5$. All data are taken at T = 2.5K, and $\mu_{0}H = 0.1T$. (a) Hall coefficient (b) Superconducting transition temperature (c) Residual resistivity, extracted from a linear fit to the resistivity above the superconducting transition. Doping on the indium (conduction electron) site versus the cerium ($f$-electron) sites have qualitatively different effects on the transport properties, despite their comparable effect on disorder scattering (i.e. residual resistivity).}
\label{fig:sup_LavsCd}
\end{figure*}

\begin{figure*}[!htbp]
\centering
\includegraphics[scale=0.8]{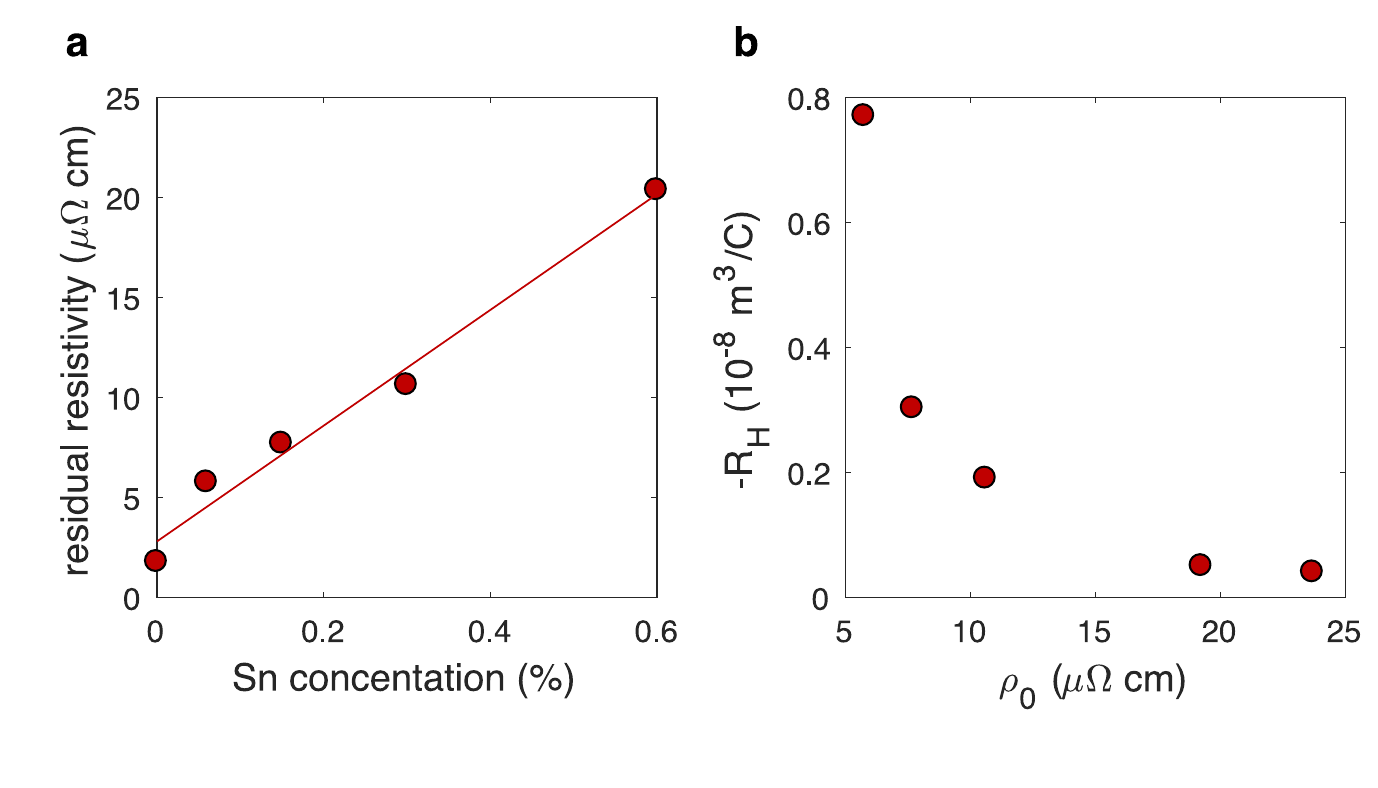}
\caption{\textbf{Effect of electron-doping on transport quantities} (a) Residual resistivity (b) Hall coefficient ($\mu_{0}H = 0.1$T; T = 5K).}
\label{fig:sup_Sn}
\end{figure*}

\begin{figure*}[!htbp]
\centering
\includegraphics[scale=0.8]{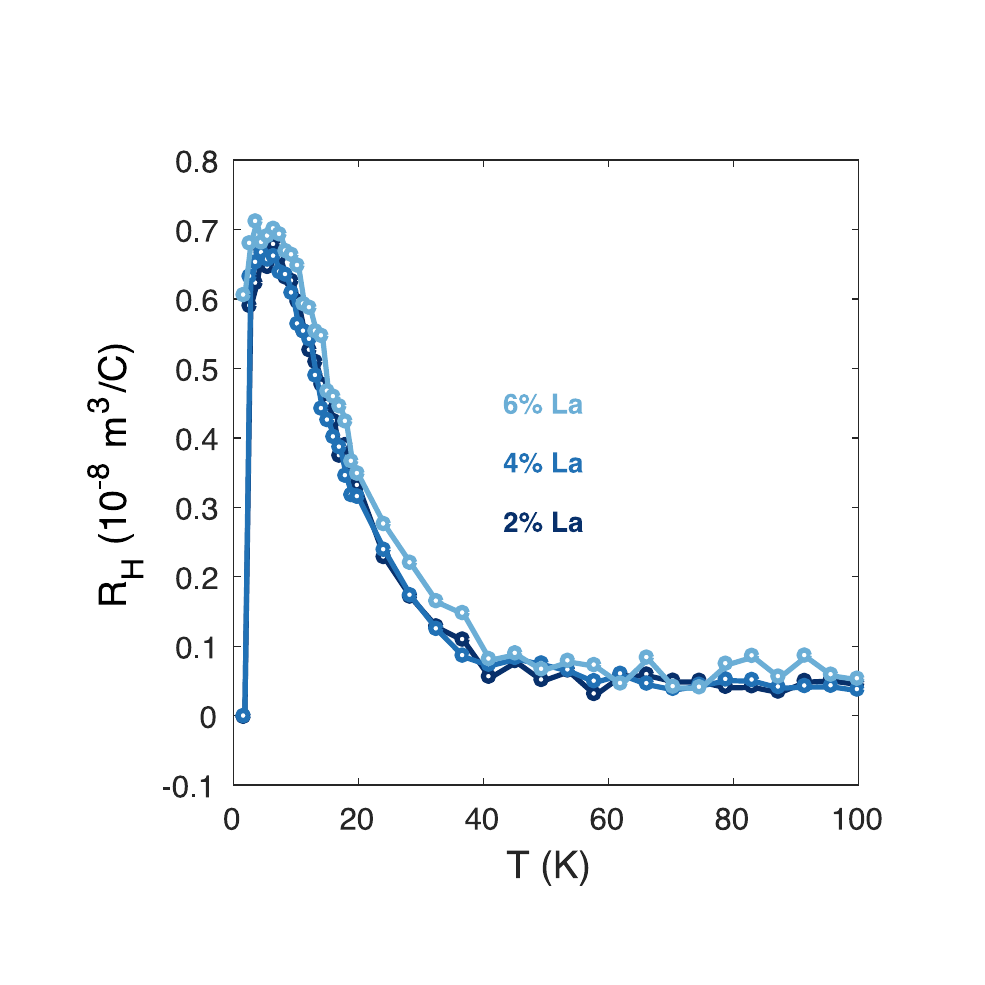}
\caption{\textbf{Hall coefficient versus temperature for lanthanum substituted samples} taken at an applied field of $\mu_{0}H = 0.1$ Tesla.}
\label{fig:sup_LavT}
\end{figure*}

Fig.~\ref{fig:sup_LavsCd} shows that the Hall coefficient is essentially unaffected by $f$-electron dilution, achieved with lanthanum substitution. On the other hand, cadmium or zinc doping of the indium site, i.e. conduction electron dilution, has a strong effect on the Hall coefficient. The effect on the Hall coefficient with Cd or Zn substitution can be plotted as a function of the induced residual resistivity scattering rate, suggesting that for a given concentration of substitutents, the effect of Cd or Zn is the same. Lanthanum doping has almost no effect on the Hall coefficient or its temperature-dependence (Fig.~\ref{fig:sup_LavT}) at the levels shown here. However, lanthanum substitution has a much stronger effect on the superconducting transition temperature than either Cd or Zn. Fig.~\ref{fig:sup_Sn} shows the analogous plots for Sn-substitution (electron-doping).

\newpage

\section*{S6 Semi-classical Boltzmann prediction of $R_{H,b}$}

In this section, we derive a semi-classical approximation for the bosonic contribution of the Hall coefficient based on the critical theory of Ref.~\cite{Senthil2004}. As in the main text, the boson and $f$-electron contribution is given by
\begin{equation}
    R_{H,b} = \sigma_{xy}^{bf}\rho_{xx}^2/(\mu_0H) \sim T^2 \sigma_{xy}^{b}/(\mu_0 H),
\end{equation}
where in the last step we substituted the observed $\rho_{xx}\sim T$. In addition due to series addition of the charged-boson and spinon resistivities, the bosons, having a much higher resistivity, dominate the transport. Therefore we take $\sigma^{bf}\approx \sigma^b$.

Following Ref.~\cite{Senthil2004}, we take the bosons to have a mass $m_b$ and charge $-e<0$, a dispersion $\epsilon = k^2/(2m_b) - \mu$, and a quartic interaction with strength $u$. The bosons are coupled to the internal gauge field enforcing the Ioffe-Larkin gauge condition. The chemical potential $\mu$ is the tuning parameter in the transition at $T=0$ with $\mu=0$ corresponding to the critical point. For comparison with the experiment, we expect that $\mu$ is proportional to the doping, $x$. We solve a semi-classical Boltzmann equation in the relaxation-time approximation for the bosons while holding $\mu$ fixed as in Ref.~\cite{Senthil2004}. The number of bosons is consequently {\it not} constant, as would be expected since the term in the Lagrangian leading to hybridization of the $f$-spinon and $c$-electron also leads to the exchange of $c$-electrons, fermionic spinons and bosons ($c\leftrightarrow f+b$).  

The scattering time has two contributions: scattering off of impurities (the dominant source of which is the dopants) and scattering off of low-energy gauge fluctuations. The impurity scattering relaxation time is given by the usual expression $\tau_i^{-1} = n_i v(k)\sigma(k)$ for velocity $v(k)$ scattering cross-section $\sigma(k)$ and impurity concentration $n_i=|\delta|$. We will assume, for simplicity, that $\tau_i^{-1}=n_i K_1 = K |\mu|$ for constants $K_1$ and $K$. The gauge-field scattering has relaxation time $\tau_g = A\beta^{3/2}/\sqrt{\beta k^2/(2m_b)}$ as in Ref.~\cite{Senthil2004}. Since the two scattering mechanisms are independent, the total scattering is given by $\tau^{-1} = \tau_{i}^{-1} + \tau_g^{-1}$. 

 After straight-forward and standard analysis, we arrive at $(\hbar = k_B =c=1)$
\begin{equation}
    \sigma_{xy}^b=-\frac{A^2\omega_ce^2(2m_b)^{3/2}}{6\pi^2m_b} \beta^{3/2}\int_0^{\infty}y^{3/2}dy\left[\frac{e^{y+a}}{(e^{y+a}-1)^2}\frac{1}{(\sqrt{y} + C |\mu|\beta^{3/2})^2+\omega_c^2A^2\beta^3}\right]
\end{equation}
where $y = \beta k^2/(2m_b), a = -\mu \beta +\beta \Sigma_b(0,0),$ and $\omega_c = e\mu_0 H/m_b$, and $C=AK$. The self-energy, $\Sigma_b(0,0)$ is given by
\begin{equation}
    \Sigma_b(0,0) =\frac{u (2m_bT)^{3/2}}{2\pi^2} \int_0^\infty \sqrt{y}dy\left[ \frac{1}{e^{y-\beta\mu}-1} - \frac{1}{y-\beta\mu} + \frac{1}{y} \right].
\end{equation}

This calculation will certainly break down at the temperature scale where $\mu>0$ and $\Sigma_b(0,0) \le \mu$. In a conventional bose liquid this would signal a transition to a superfluid phase of the bosons, which is precluded in our system if the compact $U(1)$ gauge field fluctuations are taken into account. Nevertheless the temperature scale at which $\Sigma_b(0,0) \le \mu$ still represents a crossover scale below which the boson resistivity is expected to drop sharply, thus our approximations are not valid below that scale. For the small $|\mu|$ we consider below, that crossover temperature is an order of magnitude below the peak of the graph.

\begin{figure*}[!htbp]
\centering
    \includegraphics[width=.5\textwidth]{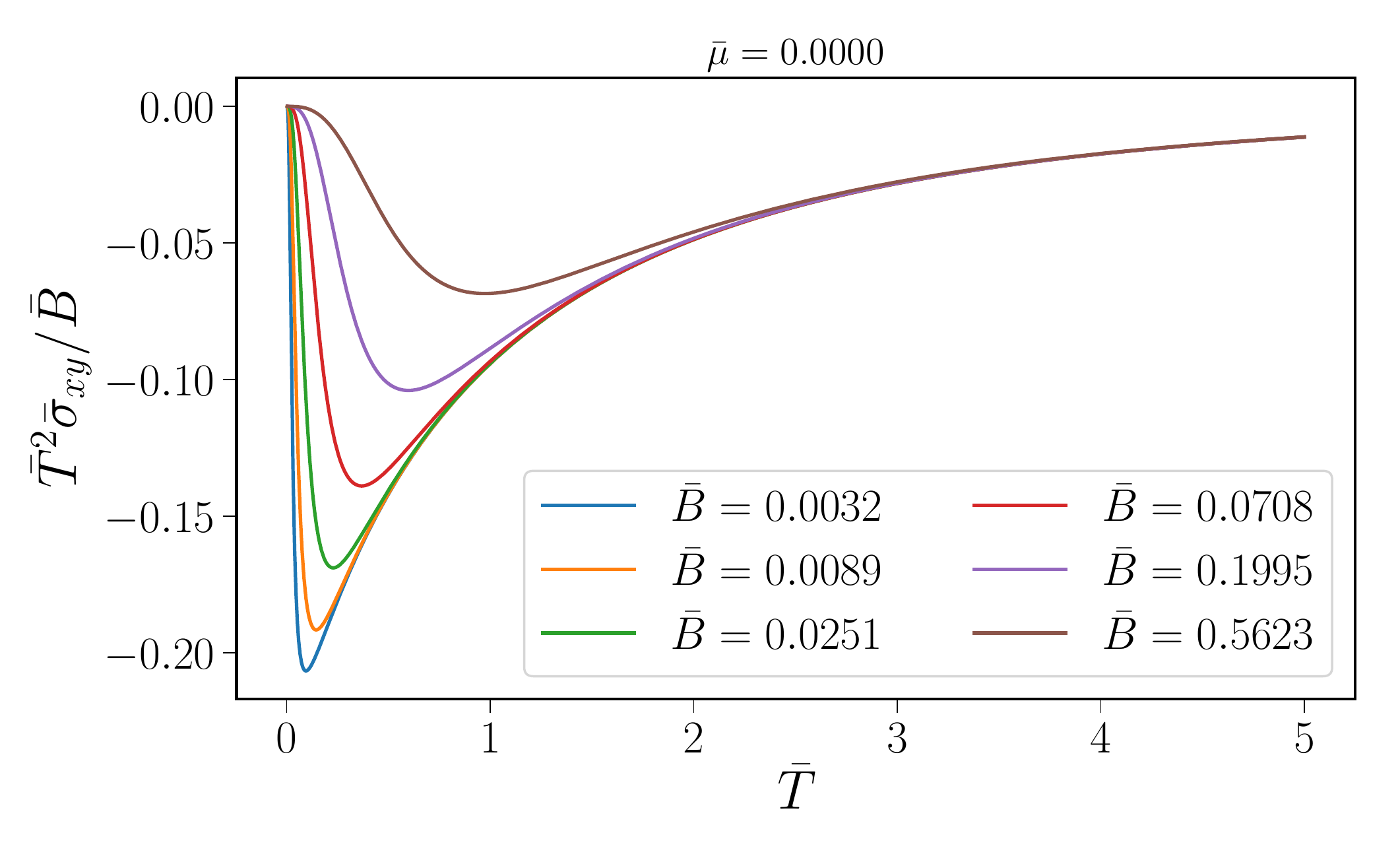}\includegraphics[width=.5\textwidth]{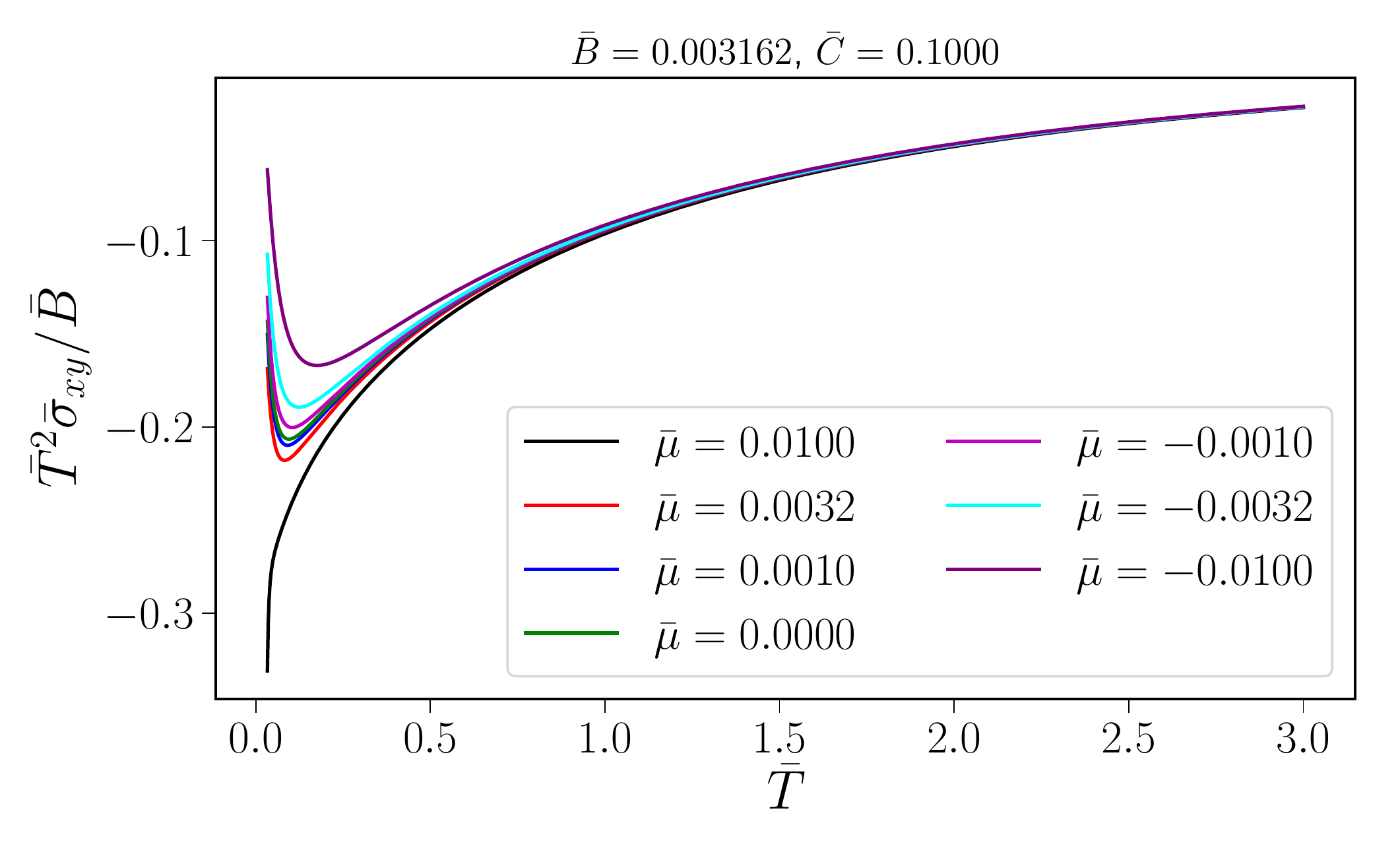}
    \caption{{\bf $R_{H,b}$ as a function of $\bar B$ and $\mu$ at small $\bar C$} We plot $\bar T^2 \bar \sigma_{xy}/\bar B \propto R_{H,b}$ as a function of $\bar T$. (a) The $\bar B$ dependence is shown at $\mu=0$, and the qualitative feature of the peak lowering and moving to higher temperature for increasing $\bar B$ is seen. The value of $\bar C$ does not matter for this plot. (b) For small $\bar B$ and $\bar C$, we see that $\mu <0$ follows the correct qualitative behavior of decreasing height of the peak and increasing the temperature of the peak, but $\mu > 0$ does not for low $\bar C$.  The divergence at $\bar \mu=0.01$ occurs because the calculation breaks down when $T \to T_c$, the superfluid transition temperature, determined by $\Sigma_b(0,0)=\mu$. At higher $\bar C$ as in figure 6S, we see that the behaviors are more similar between $\mu>0$ and $\mu<0$. }
    \label{fig:sup_theory1}
\end{figure*}

\begin{figure*}[!htbp]
\centering
    \includegraphics[width=.5\textwidth]{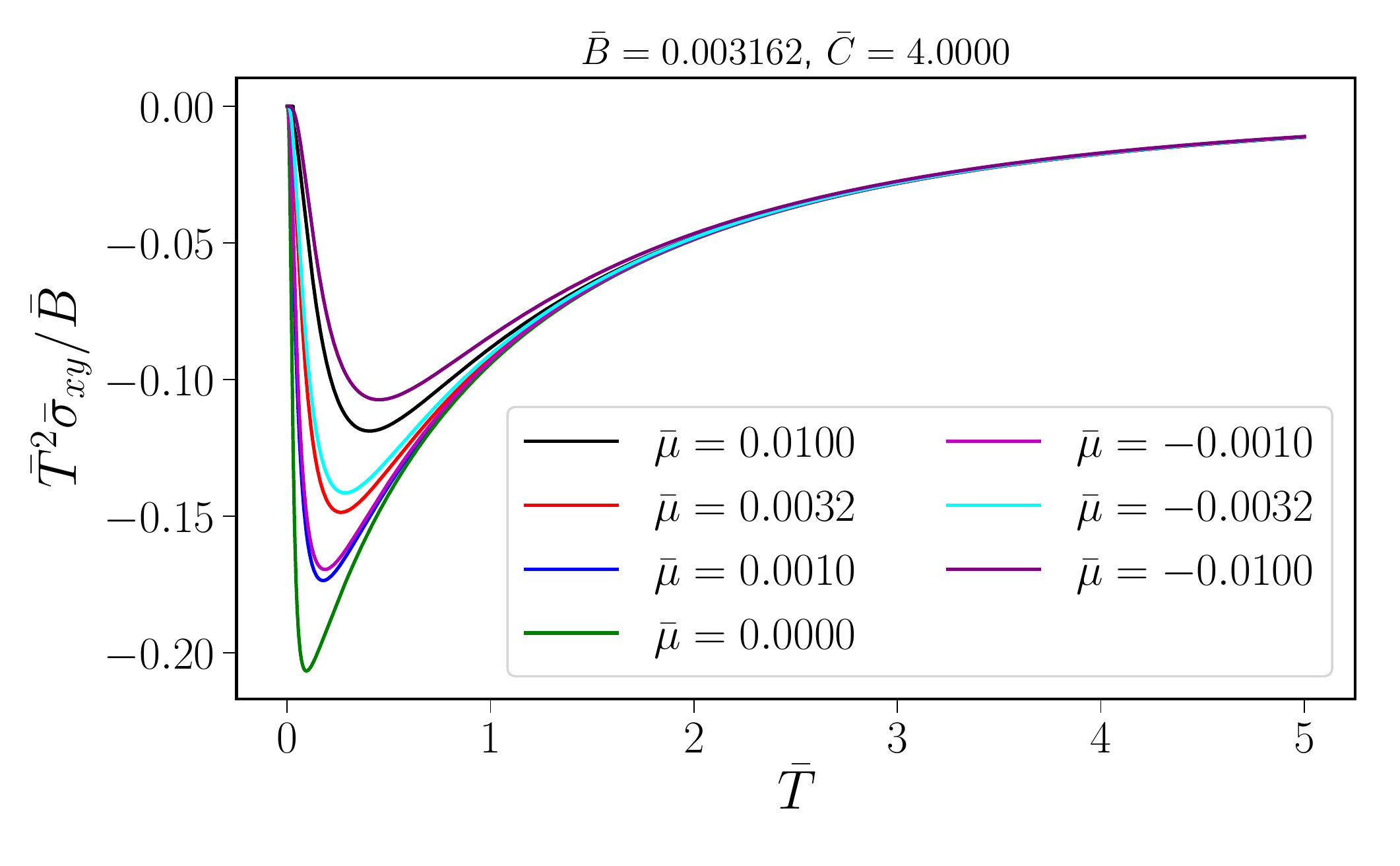}\includegraphics[width=.5\textwidth]{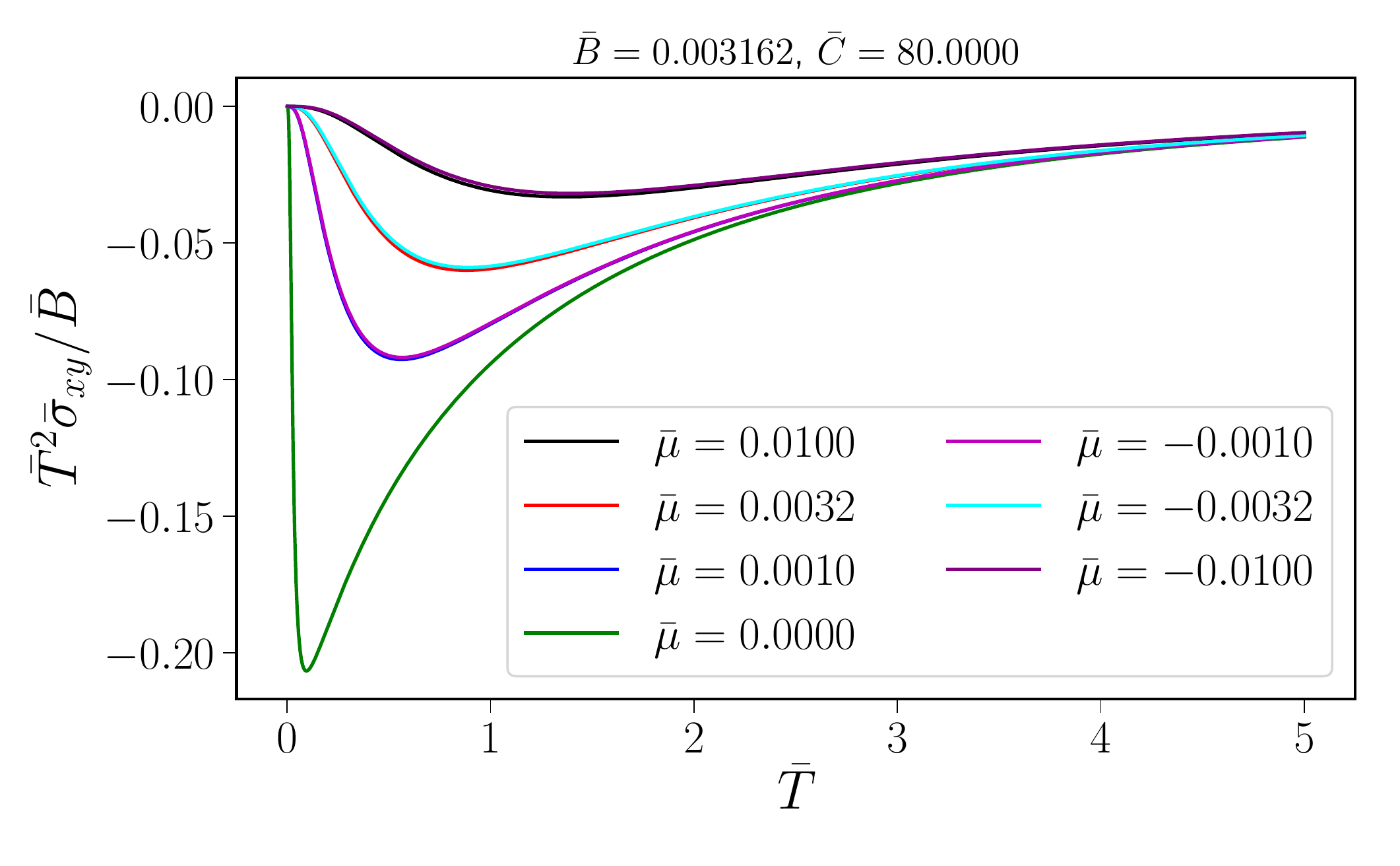}
    \caption{{\bf $R_{H,b}$ for larger $\bar C$} We plot $\bar T^2 \bar \sigma_{xy}/\bar B \propto R_{H,b}$ as a function of $\bar T$ with small $\bar B$ but larger $\bar C$. For these larger values of $\bar C$, we see that the positive and negative $\mu$ curves look more symmetric and follow the qualitative features of the experiment. The positive $\mu$ graphs only are valid for $T>T_c$ where the bosons would naively condense. For the largest $\bar \mu$ shown, this occurs at $\bar T \approx 0.03$. }
    \label{fig:sup_theory2}
\end{figure*}

We fix $1=(\bar u)^{-2}=(u(2m_b)^{3/2}/(2\pi^2))^{-2}$ as setting our energy scale, and we switch to dimensionless parameters: $\bar T = \bar u^2 T, \bar \mu = \bar u^2 \mu, \bar C = \bar u C, \bar B = \omega_c A \bar u^3\propto \mu_0 H$, and $\bar \sigma_{xy} = \sigma_{xy}^b/\mathcal C$ with $\mathcal C = A e^2(2m_b)^{3/2}/(6\pi^2m_b)$. We numerically evaluate $\bar \sigma_{xy} \bar T^2/\bar B$ vs. $\bar T$ for several choices of parameters $\bar B$, $\bar \mu$, and $\bar C$, and show some plots in Figs.~\ref{fig:fig3_diverging},~\ref{fig:sup_theory1}, and~\ref{fig:sup_theory2}.

We notice that the graphs qualitatively capture the critical curves. The temperature of the peak increases and the peak decreases with increasing $|\mu|$ (i.e. increasing doping) or increasing $H$. The asymmetry of the experiment, where the peak height decays faster for smaller electron doping than hole doping, could be explained by a difference in the value of $\bar C$ coming from a difference in scattering off of impurities.

For the critical curve, we can easily evaluate the limiting behavior for large and small $T$. In the $a \ll 1$ limit, the integrand is dominated by $y \ll 1$. In the $a \gg 1$ limit, we can approximate $e^{y+a}-1 \approx e^{y+a}$. We find,
\begin{equation}
    \lim_{H\to0}-\frac{\bar \sigma_{xy}(\mu=0)}{\bar B} = \frac{1}{\bar T^{3/2}}\int_0^{\infty} dy y^{1/2} \frac{e^{y+a}}{(e^{y+a}-1)^2}\sim
    \begin{cases}
    \frac{1}{\sqrt{a}\bar T^{3/2}} & \text{if $a \ll 1$} \\
    \frac{1}{e^a \bar T^{3/2}} & \text{ if $a\gg 1$}
    \end{cases}.
\end{equation}
Since $a(\mu=0) = \bar T^{1/2}\zeta$ for $\zeta$ an order 1 constant, we see that $|\bar \sigma_{xy}| \sim \bar T^{-7/4}$ at low temperatures and $|\bar \sigma_{xy}| \sim e^{-\zeta \sqrt{\bar T}}$ at high temperatures. As we move to $\mu<0$ and $H\to 0$, the divergence of $\bar \sigma_{xy}$ at low $\bar T$ will be cut off. Assuming as above that $\rho_{xx} \sim \bar T$, it then follows that  $R_{H,b}\sim\lim_{H\to0}\bar \sigma_{xy}(\mu=0)\rho_{xx}^2/\bar B$ does not diverge as $\bar T\to 0$. Instead, the curve has a peak structure as the $R_{H,b}$ interpolates between the $\bar T^{1/4}$ behavior at low $\bar T$ and the $e^{-\zeta\sqrt{\bar T}}$ behavior at high $\bar T$. If on the other hand we assume $\rho_{xx}=\rho_0+D T$ with a finite zero temperature resistivity $\rho_0$, then $R_{H,b}$ will diverge as $T\to 0$ at the critical point $\mu=0$.

\newpage
\section*{S7 Hall effect in LaCoIn$_5$}
\begin{figure*}[!htbp]
\centering
\includegraphics[scale=0.8]{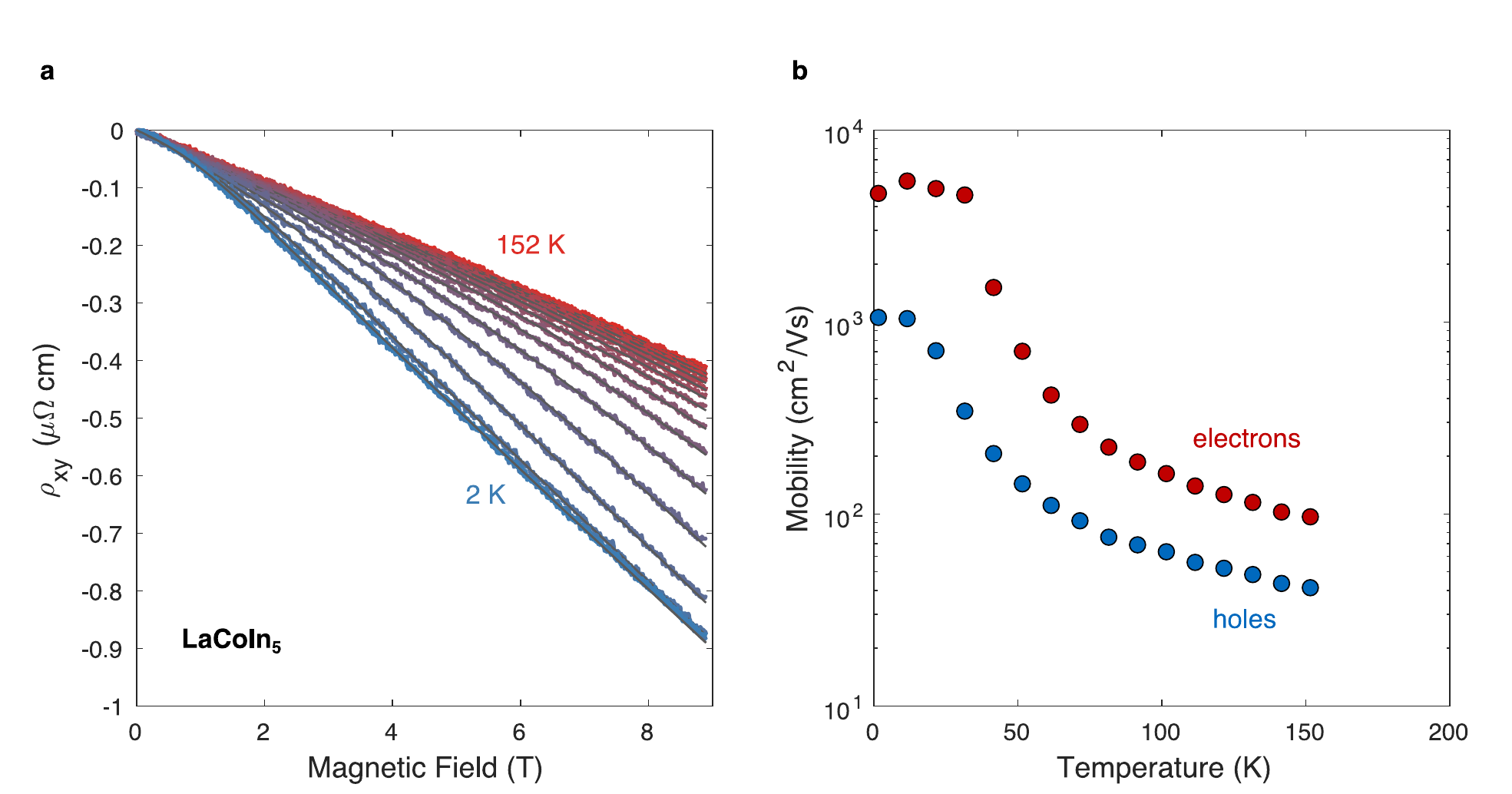}
\caption{\textbf{Hall effect in LaCoIn$_5$} (a) Hall resistivity as a function of field between T = 2K and T = 152K in 10K increments. The traces are well fit by a two-band model (grey solid lines) with an electron-like and hole-like contribution. (b) Extracted mobilities of the electron and hole bands versus temperature.}
\label{fig:sup_lacoin}
\end{figure*}

LaCoIn$_5$ is isostructural to CeCoIn$_5$, but La is missing the single $f$ level valence electron of Ce. Thus, carrier density measurements of LaCoIn$_5$ can be used to estimate the expected carrier density of the conduction bands of CeCoIn$_5$ without the $f$-electrons. The field-dependence of $\rho_{xy}$ in LaCoIn$_5$ can be fully understood with conventional transport theory. This data also serves as a useful example for discussing nonlinear Hall effect in general, and why the high-field limiting Hall coefficient gives a measure of the net carrier density. Fig.~\ref{fig:sup_lacoin}a shows isothermal field-sweeps of $\rho_{xy}$ at different temperatures for a sample of LaCoIn$_5$. We note that at the lowest temperature, the Hall coefficient approaches a field-independent constant equal to -0.07$\times 10^{-8} m^{3}/C$ at high fields. This high-field limit can be used to determine the net carrier density of this material using the standard formula $R_{H}(H\rightarrow \infty) = -\frac{1}{e(n_{h}-n_{e})}$~\cite{Pippard2009} corresponding to about 1.4 electrons per unit cell.

This material is known to have both electron-like and hole-like carriers~\cite{Nakajima2007}. We can self-consistently determine the carrier density from a multi-band fit to the Hall resistivity using the standard formula. 
\begin{equation*}
    \rho_{xy} = \frac{B}{e}\frac{(n_{h}\mu_{h}^{2} - n_{e}\mu_{e}^{2}) + (n_{h} - n_{e})\mu_{h}^{2}\mu_{e}^{2}B^{2}}{(n_{h}\mu_{h} + n_{e}\mu_{e})^{2} + (n_{h}-n_{e})^{2}\mu_{h}^{2}\mu_{e}^{2}B^{2}}
\end{equation*}
$n_{e/h}$ and $\mu_{e/h}$ are the carrier density and mobility respectively of the electron/hole band. The carrier densities are fixed, while the mobilities may vary as a function of temperature. Note that, taking the limit $H\rightarrow \infty$ of the above formula yields $\rho_{xy} = - \frac{B}{e}\frac{1}{n_{h}-n_{e}}$. The extracted carrier densities from fitting the temperature-dependent data to the above formula are found to be about $n_{e}$ = 1.42 $\times$ 10$^{22}$/cm$^{3}$ and $n_{h}$ = 0.53 $\times$ 10$^{22}$/cm$^{3}$. This gives a net carrier density corresponding to about 1.4 electrons per unit cell, consistent with the value extracted from the high-field limiting Hall coefficient at the lowest temperature. The electron and hole mobilities resulting from the fits are plotted in Fig.~\ref{fig:sup_lacoin}b.

\newpage
\section*{S8 Additional quantum oscillation measurements}

\begin{figure*}[!htbp]
\centering
\includegraphics[scale=0.8]{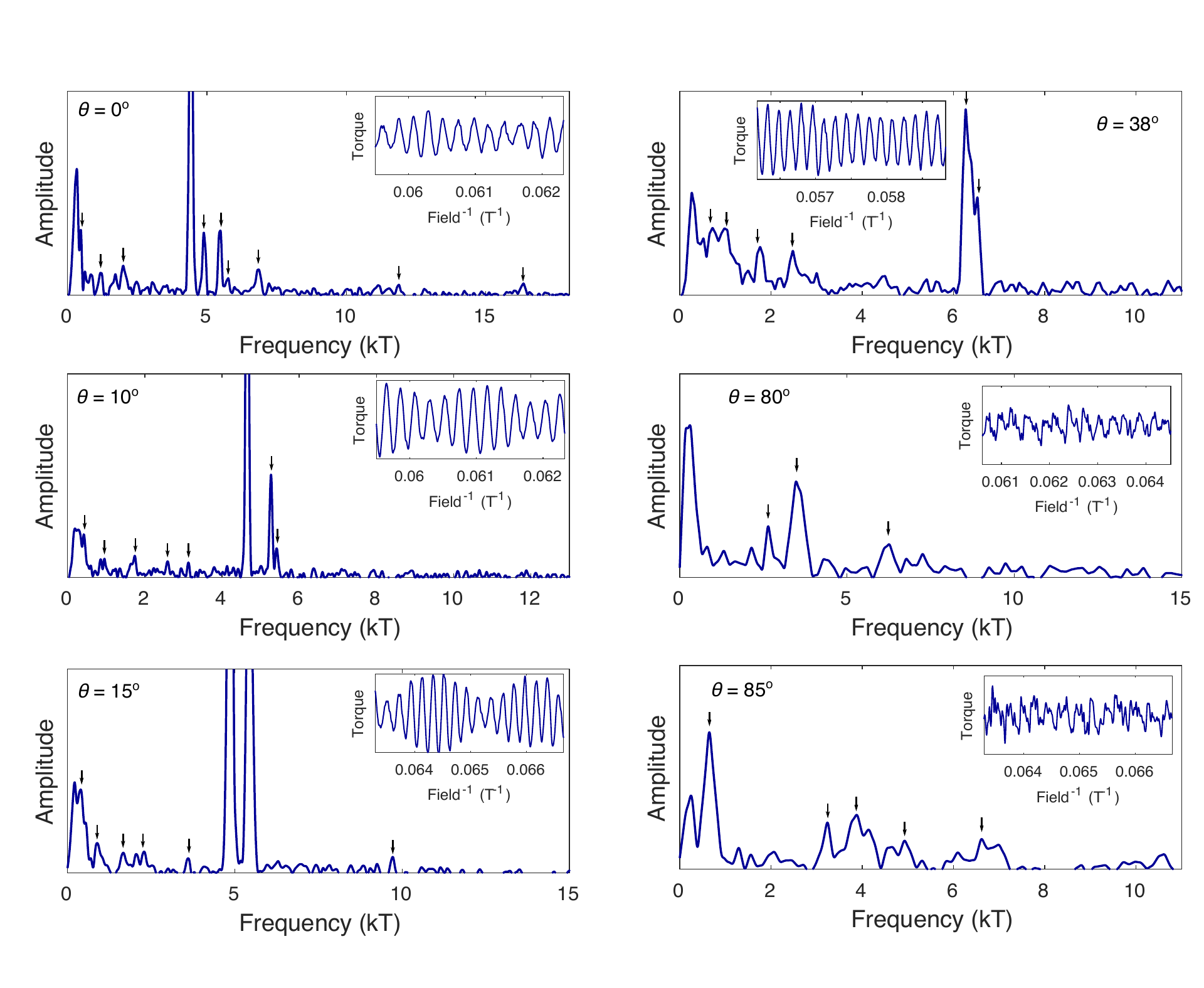}
\caption{\textbf{dHvA oscillations and spectra at different field angles for 0.9\% Sn-substituted CeCoIn$_5$} Spectra were taken in a field window of 14-17.5T. Insets show raw oscillations in the background-subtracted torque signal. Arrows mark the location of spectral peaks.}
\label{fig:sup_QO_angles}
\end{figure*}

To build the angle-dependent dHvA map shown in the main text, a combination of spectra taken over different field windows was used. Fig.~\ref{fig:sup_QO_angles} shows characteristic dHvA spectra for angles of magnetic field between [001] ($\theta = 0^{o}$) and [100] ($\theta = 90^{o}$). The spectra were taken over relatively narrow regions of magnetic field at the highest field ranges in order to resolve relatively high-frequency quantum oscillations.

\begin{figure*}[!htbp]
\centering
\includegraphics[scale=0.8]{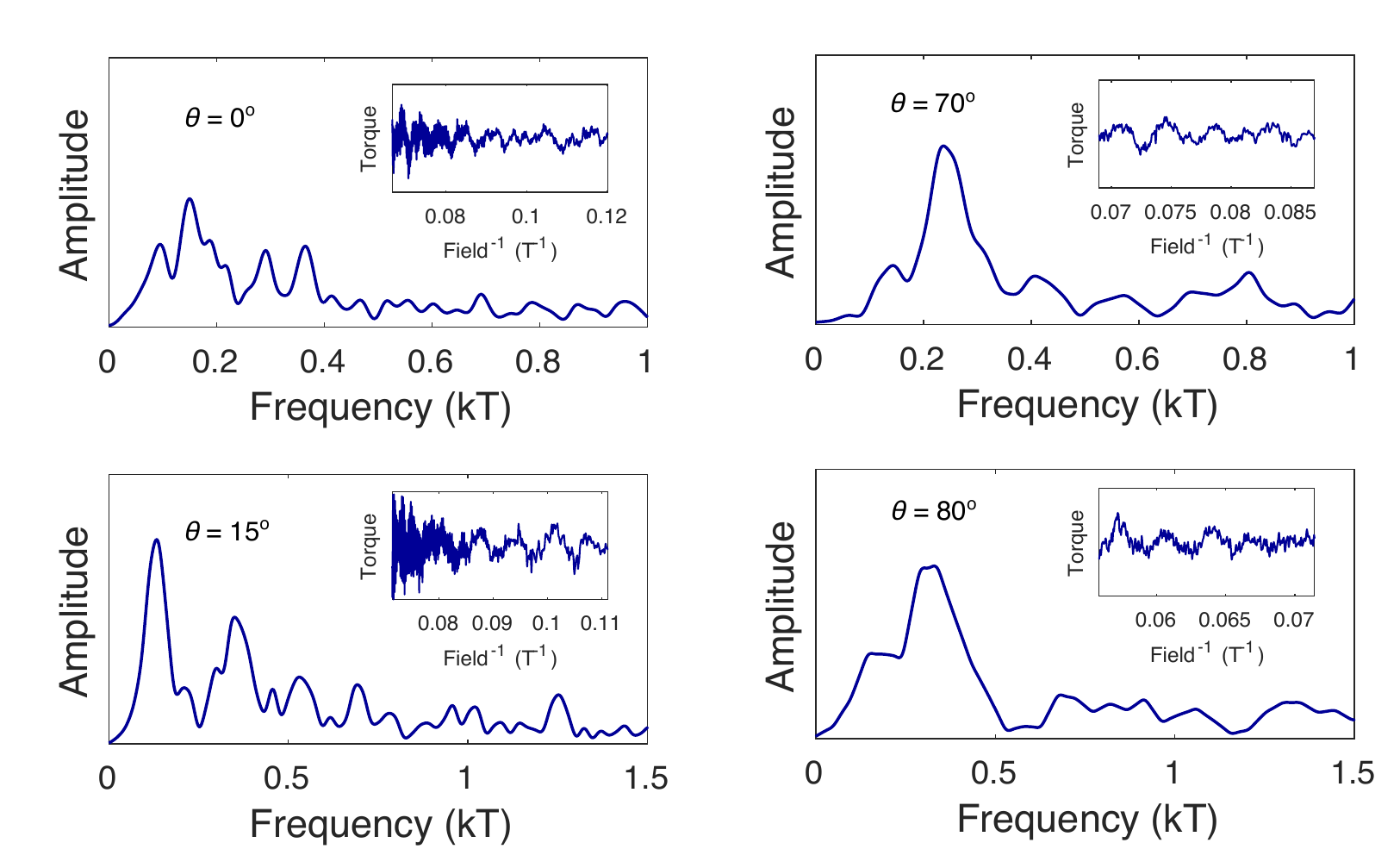}
\caption{\textbf{Low-frequency dHvA spectra and spectra at different angles 0.9\% Sn-substituted CeCoIn$_5$} Low-frequency oscillations in the signal are observed in the background-subtracted torque signal. The data were averaged on up and down magnetic field sweeps. Spectra were taken in a field window of 10-17.8T. The amplitude of these oscillations is relatively small because torque magnetometry is considerably less sensitive to quantum oscillations on isotropic Fermi surfaces.}
\label{fig:sup_QO_lowfreq}
\end{figure*}

Because there may be a number of closely spaced low-frequency orbits, the resolution of low-frequency orbits generally requires spectra taken over the full field range. In Fig.~\ref{fig:sup_QO_lowfreq}, spectra for the full field-range are shown with accompanying oscillations in the background-substracted torque signal.

\begin{figure*}[!htbp]
\centering
\includegraphics[scale=0.8]{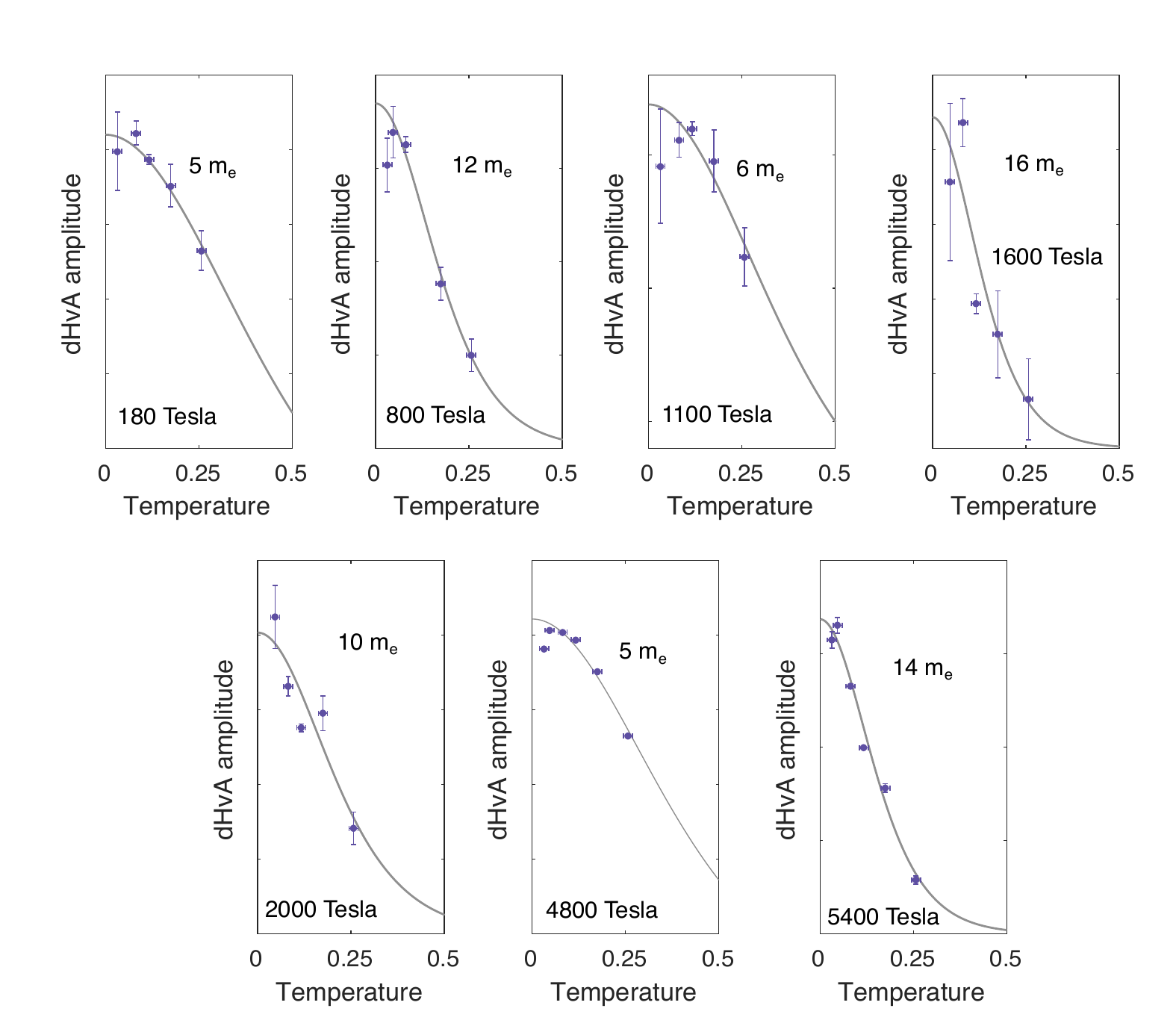}
\caption{\textbf{Temperature-dependence of dHvA oscillations and effective masses} Extracted with field 10$^{o}$ away from [001]. Each panel is labeled by the frequency of the quantum oscillation, and the effective mass extracted from the Lifshitz-Kosevich fit.}
\label{fig:sup_QO_masses}
\end{figure*}

In this section, we also extract effective masses for the $\alpha$ and $\gamma$ Fermi surface sheets observed in Fig.~\ref{fig:QO} of the main text. Fig. S8.2 shows the temperature-dependent dHvA amplitudes of each fundamental frequency. The grey line is a fit to the Lifshitz-Kosevich equation, including a constant offset fit parameter to account for imperfect background subtraction.
\begin{equation*}
    R_{T} = \frac{x}{\text{sinh}(x)}
\end{equation*}
where $R_{T}$ is the dHvA amplitude, and
\begin{equation*}
    x = \frac{2 \pi^{2} k_{B}m_{e} \mu T}{B\hbar e}
\end{equation*}
where $\mu$ is the cyclotron effective mass.

We note that the Lifshitz-Kosevich amplitude is non-monotonic in temperature in the several of the observed orbits. This deviation from Lifshitz-Kosevich behavior has been observed before in pristine CeCoIn$_5$, and attributed to spin-dependent effective masses~\cite{McCollam2005}. However, the origin of the non-Fermi liquid temperature-dependence of the dHvA orbits in this material is not agreed upon. Therefore, dHvA effective masses quoted for this material using the Lifshitz-Kosevich formula may be inaccurate. 

\begin{figure*}[!htbp]
\centering
\includegraphics[scale=0.8]{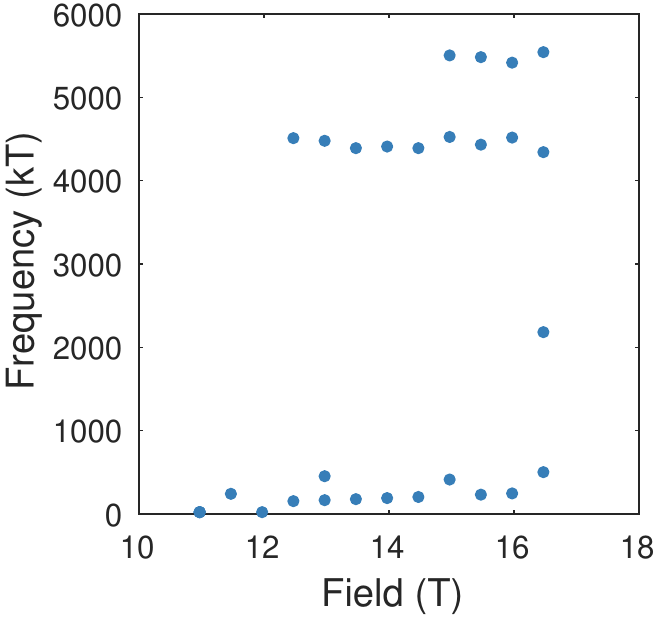}
\caption{\textbf{Dominant spectral peaks over different field windows for H$\parallel$[001]} The dots indicate dominant peaks in the spectrum for each field window.}
\label{fig:sup_QO_field}
\end{figure*}

The dominant extremal orbits of the $\alpha$ Fermi surface measured by dHvA are unchanged by the size of the magnetic field within experimental resolution (Fig.~\ref{fig:sup_QO_field}). This indicates that up to 18 Tesla, the volume of these Fermi surfaces is not affected by magnetic field. 

\newpage
\section*{S9 Hall effect measurements of CeCoIn$_5$ below 2.5 Kelvin}
\begin{figure*}[!htbp]
\centering
\includegraphics[scale=0.75]{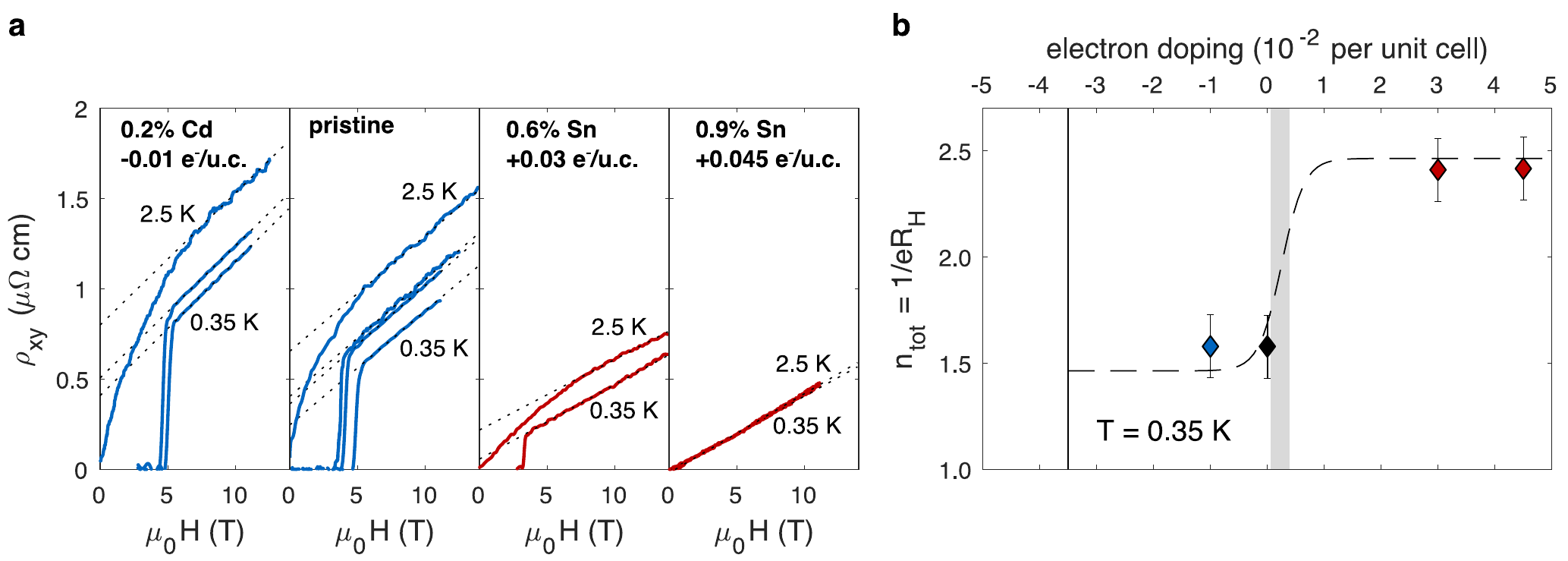}
\caption{\textbf{Hall effect measurements of CeCoIn$_5$ below 2.5 K} (a) Isothermal traces of the Hall resistivity versus magnetic field for samples with different levels of hole or electron doping. A linear fit to the high-field regime is shown by the black dotted lines. (b) Carrier density extracted from the high-field slope of the Hall resistance at 0.35 K. The data at this temperature agree well with the data at 2.5 K shown in Fig.~\ref{fig:fig1}a,b of the main text.}
\label{fig:sup_He3}
\end{figure*}

Here we present measurements of the Hall resistivity below 2.5 K. All of the data shown in the main text and Fig.~\ref{fig:fig1} are taken above 2 K. It is important to verify that the high-field limit of the Hall resistivity has been reached at 2.5 K by comparing the results to those at lower temperatures. In Fig.~\ref{fig:sup_He3}, we present data down to 0.35 K. We note that the slope of the Hall resistivity at high-fields is relatively temperature-independent below 2.5 K. Notably, the intercept of the linear fit monotonically decreases as the temperature is lowered, suggesting that the low-field nonlinear Hall effect is strongly suppressed by decreasing temperature. By contrast, the high-field slope is independent of temperature. This is a good indication that the slope of the high-field Hall resistivity at 2.5 K is representative of the high-field limiting Hall resistivity.
\end{document}